\documentclass[9pt,twocolumn,twoside,lineno]{pnas-new}
\templatetype{pnasresearcharticle}

\usepackage{graphicx}
\usepackage{bm}
\usepackage{amssymb}
\usepackage{color}
\usepackage{gensymb}

\begin{document}


\title{Signal in the noise: temporal variation in exponentially growing populations}

\author[a,1]{Eric W. Jones}
\author[b]{Joshua Derrick}
\author[c]{Roger M. Nisbet}
\author[b,d]{Will Ludington}%
\author[a]{David A.\ Sivak}%
\affil[a]{Department of Physics, Simon Fraser University}
\affil[b]{Department of Embryology, Carnegie Institution for Science}
\affil[c]{Department of Ecology, Evolution, and Marine Biology, University of California, Santa Barbara}
\affil[d]{Department of Biology, Johns Hopkins University}

\leadauthor{Jones}

\significancestatement{
Noisy exponential population growth is common and consequential.
Food contamination, microbiome assembly, and disease are sensitive to the stochastic growth of populations that originate with a few individuals. 
In this paper we quantify this noise by utilizing the distribution of times at which a growing population reaches a threshold population size.
Through this lens of so-called {\em temporal variation}, we analyze and decompose noisy population dynamics using models of stochastic population growth and bacterial growth experiments in {\em E.~coli} and {\em S.~aureus}.
In particular, we show that statistics of temporal variation are related to population growth rate and inoculum size, {\color{black} yielding a noise-based inference method for growth rate at small population sizes.}
}

\authorcontributions{E.W.J., J.D., R.M.N., W.L., and D.A.S. designed research; E.W.J. and R.M.N. performed theoretical and computational research; J.D. performed experiments; E.W.J. analyzed data; E.W.J. wrote the paper; and E.W.J., J.D., R.M.N., W.L., and D.A.S. edited the paper.}
\authordeclaration{The authors declare no competing interest.}
\correspondingauthor{\textsuperscript{1}To whom correspondence should be addressed. E-mail: eric\_jones\_2\@sfu.ca}



%



\begin{abstract}
In exponential population growth, variability in the timing of individual division events and environmental factors (including stochastic inoculation) compound to produce variable growth trajectories.
In several stochastic models of exponential growth we show power-law relationships that relate variability in the time required to reach a threshold population size to growth rate and inoculum size.
Population-growth experiments in {\em E.~coli} and {\em S.~aureus} with inoculum sizes ranging between 1 and 100 are consistent with these relationships.
We quantify how noise accumulates over time, finding that it encodes---and can be used to deduce---information about the early growth rate of a population.
\end{abstract}

\maketitle
\thispagestyle{firststyle}
\ifthenelse{\boolean{shortarticle}}{\ifthenelse{\boolean{singlecolumn}}{\abscontentformatted}{\abscontent}}{}

\firstpage[14]{4}

\dropcap{B}acteria divide, viruses replicate, and yeast cells bud, leading (if unimpeded) to exponential growth. 
Since division events are generally not evenly separated in time, even identically prepared systems will give rise to variable growth trajectories. 
Unconstrained environmental factors like stochastic inoculation further amplify this variability. 
Traditionally, the study of noisy population growth has maintained a focus on population abundance, for example quantifying a population's noise by the coefficient of variation of the abundance \cite{Hsieh2006,Stukalin2013}.
{\color{black} In this paper we offer an alternative approach by \mbox{characterizing} noisy population growth in terms of a population's {{temporal} variation}, specifically the temporal standard deviation (TSD), the standard deviation of the distribution of times at which a growing population first hits a threshold number.} 
We apply stochastic models of exponential growth to relate the TSD at large thresholds to the inoculum size and growth rate, deriving power-law relationships that match direct experimental tests in {\em Escherichia coli} and {\em Staphylococcus aureus}.

The processes of bacterial growth and division have been extensively modeled \cite{Kendall1948, Powell1955, Powell1958, Waugh1974, VoitDick1983, HirschEngelberg1966, Hinshelwood1952} and empirically characterized \cite{ScottHwa2011, Pandey2020} over the past century.
Especially over the last 15 years, experiments that enable the high-throughput, long-term observation of bacteria \cite{WangJun2010, TanouchiYou2017} have advanced the fine-grained modeling of bacterial division \cite{IyerBiswas2014, IyerBiswas2014b, HoAmir2018}.
In this paper, we propose that temporal variation is a natural lens for examining and quantifying the noisy growth of replicate bacterial populations.

We first analyze two analytically tractable models of exponential growth:
(i) the simple birth process, perhaps the most basic stochastic model of exponential growth, which assumes that each individual divides according to a Poisson process; and (ii) a model in which inoculum sizes are drawn from a Poisson distribution and growth dynamics are deterministic.
Identical power-law relationships between TSD, inoculum size, and growth rate are derived for these two models.
Then, we numerically examine age-structured population-growth models that account for an organism's age.
Last, we present bacterial growth experiments that complement and empirically ground these power-law relationships, demonstrating that statistics reporting on the temporal variation provide practical biological insights. 

\begin{figure}
\centering
    \includegraphics[width=\linewidth]{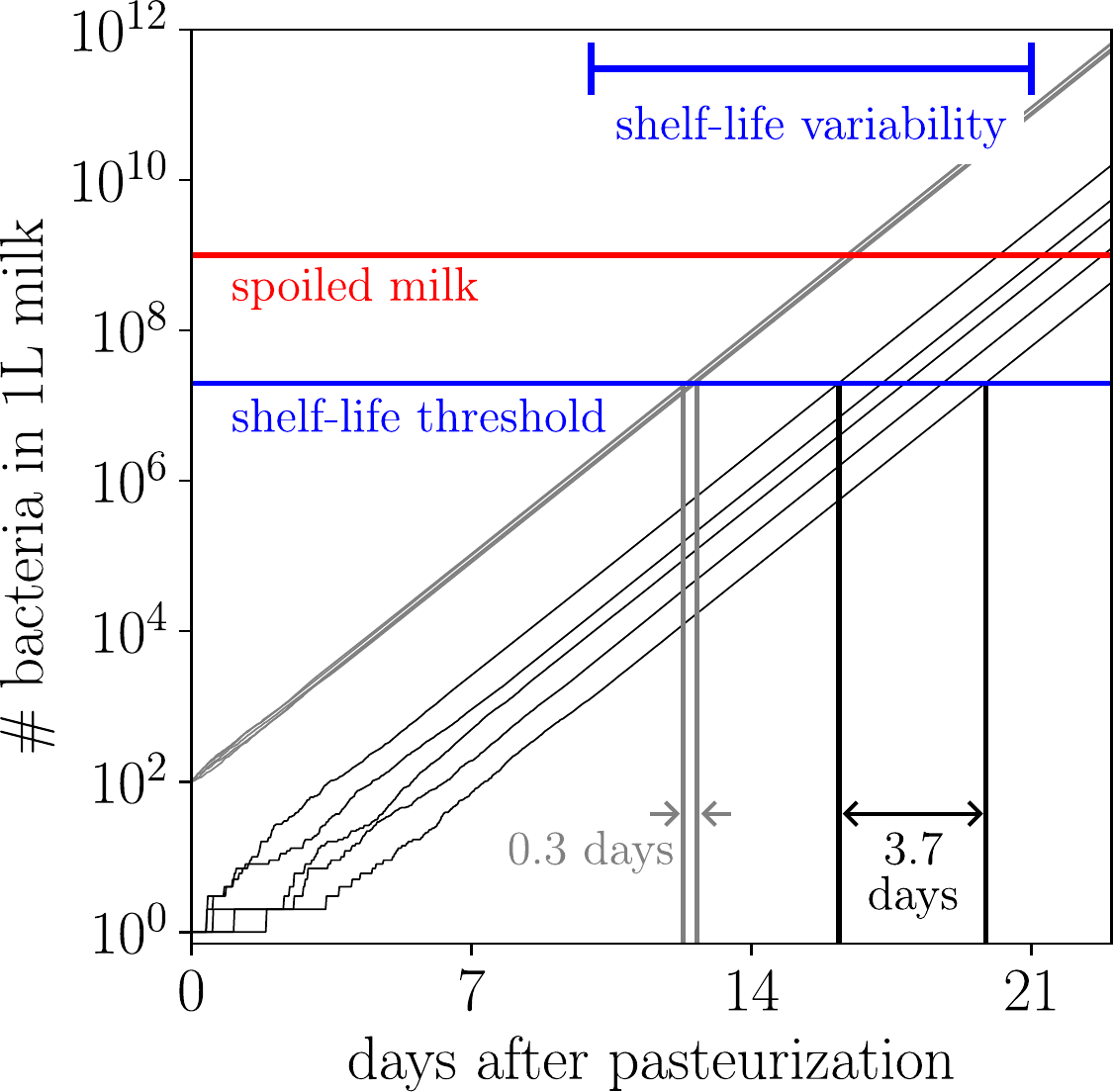}
    \caption{\textbf{Intrinsic variability contributes to the reported 11-day variation in the shelf life of milk.}
    Abundance trajectories from a simple birth process modeling the growth of {\em L.\ monocytogenes}, a common milk contaminant that divides roughly every 17 hours, inoculated with a single individual (black) or 100 individuals (gray). 
    The measured 10-21 day shelf life of milk is reported in \cite{Boor2001}.
    \label{schematic}}
\end{figure}

As a tangible example, consider milk spoilage \cite{Allen1985, Simon2001, Petrus2010}.
Milk spoilage occurs when the exponential growth of a contaminant bacteria reaches some threshold population density.
In a refrigerator at 5\textdegree C, the common bacterial contaminant {\em Listeria monocytogenes} divides every $\sim$17 hours \cite{Membre2004}.
It is straightforward to measure the distribution of times at which a number of identically prepared containers of milk spoil: if properly refrigerated, pasteurized milk has a shelf life (time to reach a bacterial concentration of 20,000 CFU/mL \cite{PasteurizedMilkOrdinance}) ranging from 10 to 21 days post processing \cite{Boor2001}. 
Figure~\ref{schematic} shows simulated abundance trajectories for the simple birth process modeling the growth of {\em L.~monocytogenes}, which indicate that a liter of milk inoculated by a single bacterium has a shelf life of roughly 17 days with a 3.7-day range, while a liter of milk inoculated by 100 bacteria has a shelf life of 12 days with a 0.3-day range. 
Nearly 4 days of the variation in the timing of milk spoilage can be accounted for by the simple birth process.
The remaining variation must be generated by other environmental factors.
Food-processing engineers that decompose noise into its \mbox{constitutive} processes might learn whether variability is \mbox{inevitable} or whether it can be mitigated.

\section*{Results}
\subsection*{Models of exponential growth}
\subsubsection*{Simple birth process}
First, consider a simple birth process in which each individual divides according to a Poisson process with rate $\mu$. 
This model was first solved in 1939 to describe the exponential growth of neutrons in nuclear fission \cite{Feller1939, Feller1968}, then subsequently used as a model of bacterial growth \cite{Kendall1948}.
This analytically tractable model permits direct calculation of statistics that report on the population's temporal variation, namely the {temporal variance} $\sigma^2_t$ and the {temporal standard deviation} $\sigma_t$ (TSD).

For a population of $n$ individuals, the probability $B_n$ per unit time that an individual will divide (conventionally the ``birth rate'' in Markov-process literature \cite{RichterDyn1972}) is
\begin{equation}
    B_n = \mu n. \label{markov}
\end{equation}
The probability $P_t(n \, | \, n_0)$ that the population consists of $n$ individuals at time $t$, given an inoculum of $n_0$ individuals, is governed by the master equation
\begin{equation}
\frac{\text{d}}{\text{d} t} P_t(n \, |\, n_0) = 
\mu (n-1) P_{t}(n-1 \, | \, n_0) - \mu n P_t(n \, | \, n_0).
\end{equation}
In $P_t(n \, | \, n_0)$, $n$ is the random variable with normalization $\sum_{n=0}^\infty P_t(n \, | \, n_0) = 1$. Using generating functions \cite{Feller1939, Kendall1948}, the solution is
\begin{align}
    P_{t}(n \, | \, n_0) &= {n-1 \choose n_0-1} 
    e^{-\mu n_0 t} (1 - e^{-\mu t})^{n-n_0},
    \label{P_n}
\end{align}
for binomial coefficient ${i \choose j} \equiv i!/j! (i-j)!$.
The first two cumulants are the average abundance
\begin{equation}
    \langle n \rangle
    = n_0 e^{\mu t}, \label{mean_abundance}
\end{equation}
which grows exponentially, and the variance
\begin{equation}
    \langle (n - \langle n \rangle )^2\rangle
    = n_0 e^{\mu t}(e^{\mu t} - 1). \label{P_var}
\end{equation}

The first-passage-time distribution $P_{\Omega}^{\text{FP}}(t \, | \, n_0)$ is the distribution of times at which a population with inoculum size $n_0$ first reaches $\Omega$ individuals \cite{Redner2001}.
Since the simple birth process yields monotonic abundance trajectories,
the reaction probability $R_{\Omega}(t \, | \, n_0)$ that at time $t$ the population size is greater than or equal to population threshold $\Omega$ is related to the first-passage-time probability $P_{\Omega}^{\text{FP}}(t \, | \, n_0)$:
\begin{align}
    R_{\Omega}(t \, | \, n_0) &= 1 - \sum_{i=n_0}^{\Omega-1} P_t(i \, | \, n_0)
    =\int_0^t P_{\Omega}^{\text{FP}}(\tau \, | \, n_0)
    \, \text{d}\tau.
\end{align}
Therefore,
\begin{align}
    P_{\Omega}^{\text{FP}}(t \, | \, n_0) = - \sum_{i=n_0}^{\Omega-1} \frac{\text{d} P_t(i
    \, | \, n_0)}{\text{d}t},
\end{align}
yielding (Supplementary Information, Section A)
\begin{align}
    P_{\Omega}^{\text{FP}}(t \, | \, n_0) 
    &= \mu(\Omega-n_0) {\Omega-1 \choose n_0 - 1} \nonumber \\
    &\quad \times (e^{-\mu t})^{n_0}(1 - e^{-\mu t})^{\Omega-n_0-1} .
    \label{fpt_dist}
\end{align}

\begin{figure}[t]
    \includegraphics[width=.48\textwidth]{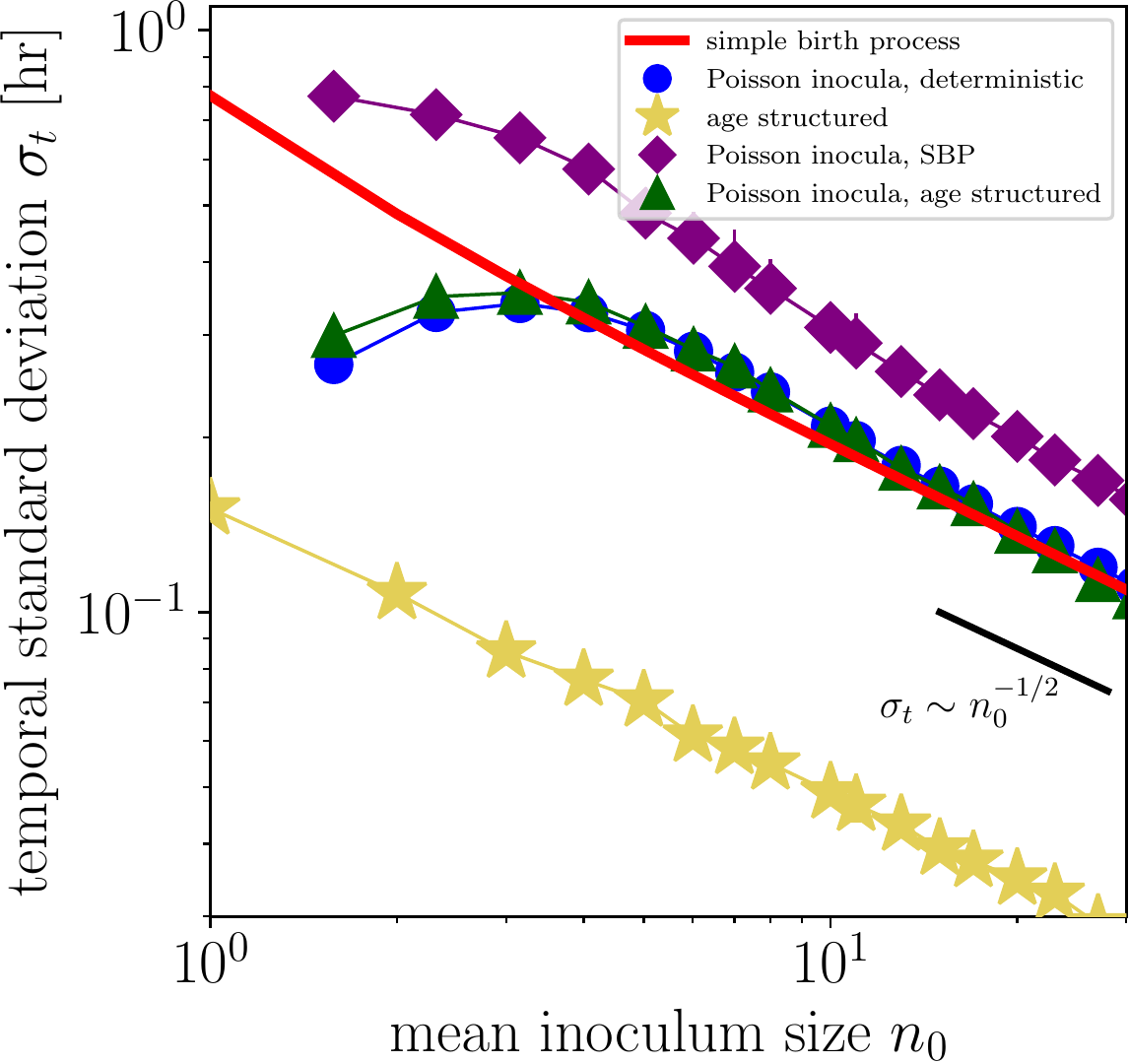}
    \caption{
    \textbf{Temporal standard deviation (TSD) scales inversely with the square root of inoculum size for five models of stochastic exponential growth.}
    For each model, inocula are either exact or Poisson-distributed, and growth either obeys the simple birth process (SBP), deterministic exponential growth, or age-structured growth. 
    The growth rate $\mu$ for the simple birth process and deterministic growth is 1.66/hr, corresponding to a 25-minute division time. 
    The division-time distribution for the age-structured population-growth model has a 25-minute mean division time and a 22\% coefficient of variation (Fig.~S1).
    At least $n=2,000$ replicates were simulated for each model and inoculum size.
    Error bars, which are typically smaller than the corresponding symbol, show 95\% confidence intervals (Methods).
    For Poisson-distributed inocula, the x-axis reports the zero-truncated mean inoculum size.
    Lines are a guide to the eye.
    \label{model_TSDs}}
\end{figure}

The mean first-passage time $\langle t \rangle_{\Omega \, | \, n_0}$ to reach threshold $\Omega$ starting from $n_0$ individuals is (SI, Section B)
\begin{subequations}
\begin{align}
    \langle t \rangle_{\Omega \, | \, n_0}
    &= \frac{1}{\mu}\left(\frac{1}{n_0} + \frac{1}{n_0 + 1} + \cdots +
    \frac{1}{\Omega-1} \right), \label{FPT_exp} \\
    &\approx \frac{\ln \Omega}{\mu} - \frac{\ln n_0}{\mu} \quad \text{for large } \Omega \gg n_0. \label{testlabel}
\end{align}
\end{subequations}
The {\em temporal variance} $\sigma^2_t \equiv \left\langle (t - \langle t\rangle )^2 \right\rangle_{\Omega \, | \, n_0}$ is (SI, Section B)
\begin{equation}
   \sigma_t^2 =
   \frac{1}{\mu^2}\left[ \frac{1}{n_0^2} + \frac{1}{(n_0+1)^2} + \cdots +
    \frac{1}{(\Omega-1)^2}\right], \label{FPT_var}
\end{equation}
and therefore the {\em temporal standard deviation} is
\begin{subequations}
\begin{align}
   \sigma_t &= \frac{1}{\mu}\left[ \frac{1}{n_0^2} + \frac{1}{(n_0+1)^2} +
    \cdots + \frac{1}{(\Omega-1)^2}\right]^{1/2}
   \label{temporal_variation} \\
   &\approx \frac{1}{\mu n_0^{1/2}} \quad \text{for large } \Omega \gg n_0.
   \label{temporal_variation_b}
\end{align}
\end{subequations}
This exact relationship between TSD, growth rate, and inoculum size for the simple birth process is plotted in Fig.~\ref{model_TSDs} (red curve).
Later we will show that this relationship is in complete accordance with bacterial population-growth experiments.

The mean first-passage time (\ref{FPT_exp}) and the temporal variance (\ref{FPT_var}) can alternatively be solved by leveraging the Markovianity of the simple birth process: a population of size $n$ experiences an exponentially distributed waiting time with mean $1/\mu n$ before an individual in the population divides, and the variance of this waiting-time distribution is $1/(\mu n)^2$.
Waiting times are independent, so moments of the first-passage-time distribution are simply the sum of the moments of the waiting-time distributions. However, this approach does not immediately provide the first-passage-time distribution Eq.~(\ref{fpt_dist}).

\subsubsection*{Poisson-distributed inocula undergoing deterministic exponential growth}
Departing from the assumption that populations are initialized with exactly $n_0$ individuals, we next consider populations with Poisson-distributed inocula that grow deterministically.
This scenario is relevant because bacterial inoculation in our experiments---performed by pipetting a fixed volume of a dilute solution of bacteria---resulted in Poisson-distributed inocula (Fig.~S2).
Populations with Poisson-distributed inocula are more variable than populations that are exactly inoculated, as variability in the inoculum size propagates through the growth dynamics.

As before, replicate populations give rise to a distribution of abundance trajectories.
We exclusively consider trajectories with nonzero inoculum sizes such that the probability $P_{n_0}(k)$ of starting with $k$ individuals is 
\begin{equation}
    {P}_{n_0}(k) = \frac{e^{-n_0} n_0^k}{k! (1 - e^{-n_0})}, \label{poisson_dist}
\end{equation}
corresponding to mean inoculum size $n_0/(1-e^{-n_0})$ for Poisson shape parameter $n_0$. 

We consider deterministic population growth
\begin{equation}
    n(t) = k e^{\mu t},
\end{equation}
a simplifying assumption that implies the abundance $n(t)$ takes on non-integer values.
The random variable
\begin{equation}
    T(M) \equiv \frac{1}{\mu} \ln(\Omega/M) \label{TSD_rv}
\end{equation}
is the first-passage time at a threshold $\Omega$ given that the inoculum size is a random variable $M$.
The temporal standard deviation can be computed exactly, albeit opaquely: 
\begin{align}
    \sqrt{\langle T(M)^2 \rangle - \langle T(M) \rangle ^2} &= \left\{ \frac{e^{-n_0}}{\mu^2 (1 - e^{-n_0})} \phantom{\left[\sum_{i}^{j}\right]^2} \right. \nonumber \\
    &\hspace{-10em} \times \left. \left[ \sum_{k=1}^\infty \frac{ (\log{k})^2 \, n_0^k}{k!} - \left( \sum_{k=1}^\infty \frac{n_0^k \, \log{k}}{k!} \right)^2 \frac{e^{-n_0}}{1-e^{-n_0}} \right] \right\}^{1/2}.
   \label{exact_poisson}
\end{align}
This temporal standard deviation is plotted as a function of mean inoculum size in Fig.~\ref{model_TSDs} (blue circles). 

To obtain the TSD at large $n_0$, first note that for large $n_0$ the Poisson distribution Eq.~(\ref{poisson_dist}) is well-approximated by a normal distribution with mean $n_0$ and variance $n_0$, and the quantity $1 - e^{-n_0}$ is well-approximated by 1.
Then, the ``delta method'' \cite{Oehlert1992, VerHoef2012} gives access to the mean and variance of the random variable $T(M)$ in terms of cumulants of $M$: 
\begin{subequations}
\begin{align}
    \langle T(M) \rangle &= T( \langle M \rangle ) 
    + \frac{T''(\langle M \rangle)}{2} \left( \langle M^2 \rangle - \langle M \rangle^2 \right) \nonumber \\
    &\quad + \text{higher-order terms} \label{delta_1}\\
    &= \frac{\log(\Omega/n_0)}{\mu} + \frac{1}{2 \mu 
    n_0} +
    O\left( \frac{1}{n_0^2} \right),
\end{align}
\end{subequations}
and
\begin{subequations}
\begin{align}
    \langle T(M)^2 \rangle - \langle T(M) \rangle^2 
    &= [T'(\langle M \rangle)]^2 \left(\langle M^2 \rangle - \langle M \rangle^2 \right) \nonumber \\
    &\quad + \text{higher-order terms} \label{delta_2}
    \\
    &= \frac{1}{\mu^2 n_0} +
    O\left( \frac{1}{n_0^2} \right),
\end{align}
\end{subequations}
where the higher-order terms depend on third and higher cumulants of $M$ that vanish when $M$ is normally distributed.

Therefore, for large $n_0$, TSD and inoculum size are related by:
\begin{align}
    \sqrt{\langle T(M)^2 \rangle - \langle T(M) \rangle^2} \approx \frac{1}{\mu
    n_0^{1/2}}. \label{poisson_scaling}
\end{align}
This is the same relationship between TSD, inoculum size, and growth rate as for the simple birth process with exact inoculation, Eq.~(\ref{temporal_variation_b}). 

\subsubsection*{Age-structured population growth \label{sec_ic}}
Organismal division is carefully choreographed, and we next turn to models that resolve some of the structure of individual division events.
We performed agent-based simulations of age-structured population growth in which division-time distributions fully describe the timing of division events (Methods).
To be precise, this model is a type of Bellman-Harris stochastic branching process \cite{BellmanHarris1948}.
We used an approximately normal division-time distribution with 25-minute mean and 22\% coefficient of variation \cite{Kendall1948}. 
Inoculated individuals were assumed to be at a random time along their division cycle.

From these simulated abundance trajectories, TSDs were evaluated at a threshold of 500 individuals and are plotted as gold stars in Fig.~\ref{model_TSDs}. While the more complicated structure of this population-growth model prevents analytic examination, the scaling of TSD with inoculum size visually follows the -1/2 power law predicted by the simple birth process and by Poisson-distributed inocula with exponential growth.

\subsubsection*{Comparing models of population growth}
Last, we simulated models for every combination of inoculation (exact or Poisson-distributed) and population growth (simple birth process, deterministic, or age-structured) (Methods).
Figure \ref{model_TSDs} shows numerically calculated TSDs for Poisson-distributed inocula obeying the simple birth process (purple diamonds), and for Poisson-distributed inocula undergoing age-structured growth (green triangles).

The models showcased in Fig.~\ref{model_TSDs} ostensibly describe the same organism, but differ in their biological assumptions about inoculation and growth.
The relationships between TSD and inoculum size quantify the effects of these assumptions on observed temporal variation.
In particular, we found that the relationship between TSD and inoculum size for a biologically faithful model that captured stochasticity in inoculation and growth (green triangles) was similar to the relationship for the simple birth process (red line).

{\color{black} 
The mean trajectories of the 
different
stochastic growth models---unlike the temporal variation---are nearly indistinguishable for a given inoculum size, highlighting an advantage of noise-based analyses.}
For example, TSDs for age-structured growth are $\sim$5 times smaller than for the simple birth process, a consequence of the fact that tighter division-time distributions give rise to less variable growth trajectories \cite{Kendall1948}.
Especially for organisms with constrained division-time distributions, the noise from Poisson inoculation dominates the noise due to growth, which explains why the blue circles and green triangles are so similar in Fig.~\ref{model_TSDs}. 
For exactly inoculated populations, broadening the age-structured division-time distribution from 22\% coefficient of variation to 100\% interpolates between the gold stars and red line; similarly, for Poisson-distributed inocula, it interpolates between the green triangles and purple diamonds. 

{\color{black} Temporal variances approximately add: the temporal variance of populations with Poisson-distributed inocula that follow the simple birth process is roughly the sum of the temporal variance of exactly inoculated populations growing according to the simple birth process and the temporal variance of populations with Poisson-distributed inocula and deterministic growth.}

We have used mathematical models of varying resolution to describe population growth, trading off biological realism for analytic tractability. 
For example, the simple birth process assumes that a bacterium's age is irrelevant to its division, but it can be solved exactly.
Going forward, we focus on the relationship Eq.~(\ref{temporal_variation}) between TSD and inoculum size for the simple birth process (red line), but emphasize that we would reach similar conclusions---at the price of analytic tractability---if we instead used the relationship for Poisson-distributed inocula and age-structured population growth (green triangles).

\subsection*{Bacterial growth experiments\label{experiment}}
To empirically test the relationship between TSD and inoculum size, we measured the growth of {\em E.\ coli} and {\em S.\ aureus}.
At least 30 biological replicates were prepared for each inoculum size and grown over one or two days.
Inoculum sizes were set by pipetting a dilute solution of bacteria growing in mid-log phase into a 96-well plate.
Spot plating the same volume of this dilute solution established mean inoculum sizes and confirmed that inoculum sizes were Poisson distributed (Fig.~S2).
Bacterial abundance was inferred by measuring the optical density of each well every 2 minutes.

\begin{figure}[t]
    \includegraphics[width=0.47\textwidth]{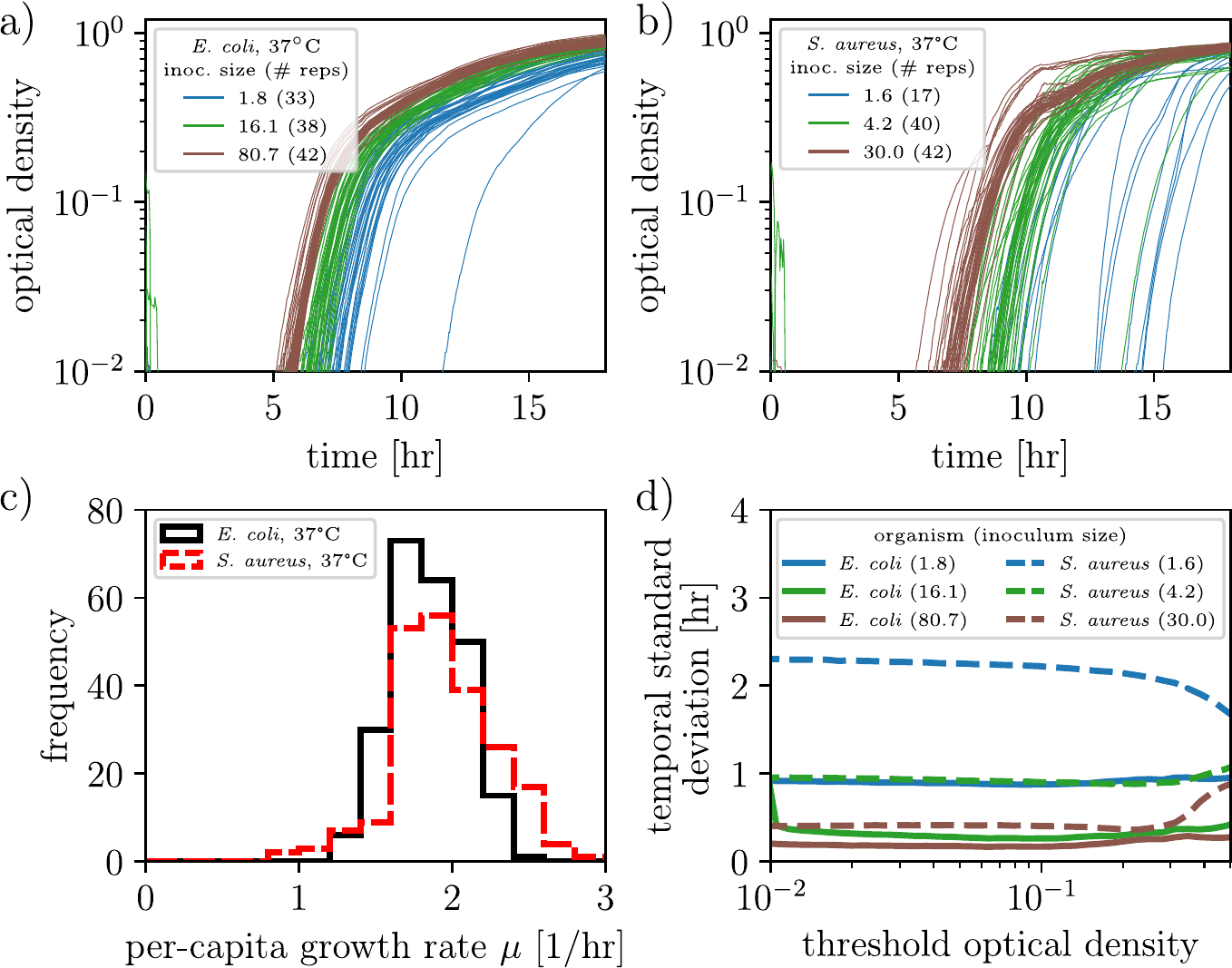}
    \caption{\textbf{Empirical analyses of bacterial growth trajectories.}
    (a, b) Measured abundance trajectories in {\em E.\ coli} and {\em S.\ aureus} as functions of time for different mean inoculum sizes.
    (c) Distribution of log-phase growth rates pooled across replicates and inoculum sizes, evaluated at an optical density of 0.03 (Methods). 
    (d) Temporal standard deviations as functions of threshold optical density for different mean inoculum sizes. 
    \label{growth_trajs}}
\end{figure}

Figures~\ref{growth_trajs}a and \ref{growth_trajs}b show representative subsets of abundance trajectories for {\em E.~coli} and {\em S.~aureus}, respectively.
Bacteria grow exponentially until they reach an optical density of $\sim$0.2, then grow more slowly until they reach carrying capacity.
During the exponential-growth phase, each individual's growth rate is $\sim$2/hour ($\sim$20-30-minute division times). 
Figure~\ref{growth_trajs}c shows the distribution of growth rates across replicates, calculated as the slope of the log-transformed optical-density time series evaluated at a threshold optical density of 0.03 (Methods).
Measuring the growth rate $\mu$ at an optical density of 0.02 increases its value by 15\%, while evaluating it at 0.05 decreases its value by 10\%.

{\color{black} Lag phase, the time period 
during which bacteria do not divide after being transferred to a new environment, could in principle affect the temporal variation of a growing population \cite{RolfeHinton2012, Bertrand2019, MorenoAckermann2020}. 
However, we expect lag phase did not significantly impact our experiments: in our setup, bacteria in log phase (exponential growth) were back-diluted into fresh and otherwise-identical media so that their growth never halts (Methods).
To check this expectation, for each inoculum size in Fig.~\ref{growth_trajs}a we calculated that the time required to reach an OD threshold of 0.03 
($\sim$$10^7$ CFUs)
assuming deterministic exponential growth with 1.8/hr growth rate and no lag phase exceeded the average empirically observed times by 30--60 minutes (Methods).
A significant lag phase, by comparison, would imply that the first-passage time for the deterministic model without lag phase is shorter than the empirically observed time.}

Equation~\eqref{temporal_variation} predicts that the temporal standard deviation for the first-passage time to threshold $\Omega$ asymptotes to a constant value for $\Omega \gtrsim 50$.
Figure~\ref{growth_trajs}d confirms this prediction: the TSD is approximately the same for threshold optical densities 0.01--0.3 {\color{black} (corresponding to millions to tens of millions of bacteria).} 

\begin{figure}[t]
    \includegraphics[width=0.47\textwidth]{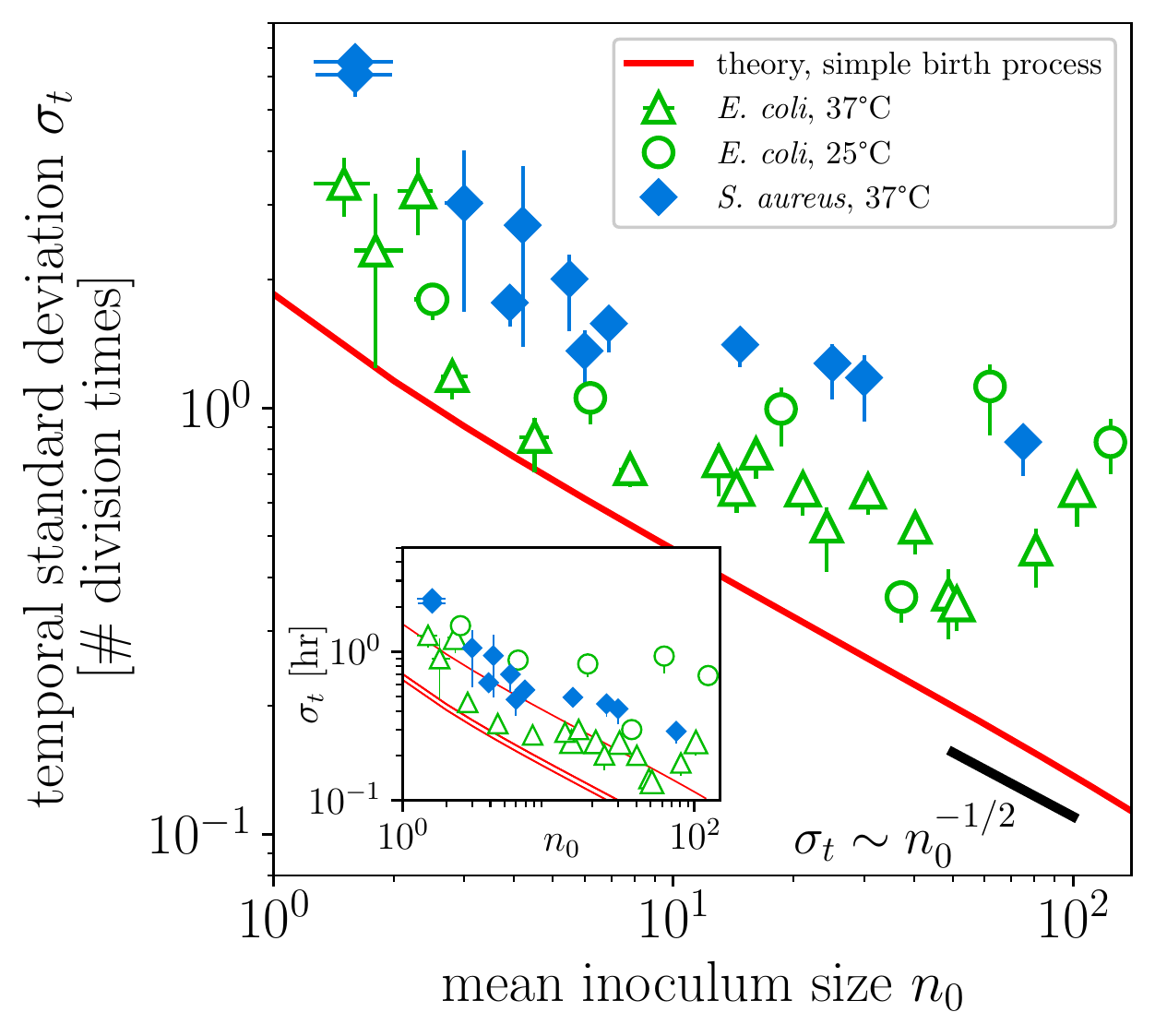}
    \caption{\textbf{Temporal standard deviation scales inversely with the square root of the inoculum size in bacterial growth experiments.}
    Temporal standard deviations for a total of 35 inoculum sizes in {\em E.~coli} and {\em S.~aureus}, in units of division times (at least 15 replicates per inoculum size, average 40).
    (inset) TSDs plotted in units of hours. 
    The theoretical TSD for a given inoculum size [red line, Eq.~(\ref{temporal_variation})] derived for the simple birth process lies under every experimental measurement (not a fit).
    Population-growth experiments were noisier than the limit of the simple birth process.
    Error bars indicate 68\% confidence intervals of the mean (Methods).
    (inset) Red lines from top to bottom calculated with growth rates from {\em E.~coli} at 25\textdegree C, {\em E.~coli} at 37\textdegree C, and {\em S.~aureus} at 37\textdegree C.
    \label{fig:power_law}}
\end{figure}

Bacterial growth experiments were performed for 35 inoculum sizes, yielding 1,381 total growth curves.
Figure~\ref{fig:power_law} shows how TSDs depend on inoculum size in units of hours (inset) and in units of division times (main figure).
An organism's division time is defined as $\ln(2)/\mu$ for growth rate $\mu$: at 37\textdegree C, {\em E.~coli} and {\em S.~aureus} have division times of $\sim$22 minutes, and at 25\textdegree C {\em E.~coli} has a division time of $\sim$50 minutes (Methods).
Presenting the data in terms of division times rather than hours collapses the TSDs of {\em E.~coli} at 25\textdegree C onto the TSDs of {\em E.~coli} at 37\textdegree C in Fig.~\ref{fig:power_law}.

In the stochastic growth models considered in Fig.~\ref{model_TSDs}, noise in abundance trajectories is generated either by variability in the timing of division events or by variability in inoculum size.
Since additional extrinsic sources of noise {\color{black} (like differing media conditions, temperature fluctuations, or lag phase)} are not included in this accounting, we hypothesized that Eq.~(\ref{temporal_variation}) would underestimate the noise in the empirical measurements.
This hypothesis is borne out by the data: in Fig.~\ref{fig:power_law} the temporal standard deviation predicted by the simple birth process (red line) lies below all 35 experimentally tested inoculum sizes (colorful symbols).
This, our main empirical result, provides strong experimental support for the relationship (\ref{temporal_variation}) as a lower bound to the temporal variation of an exponentially growing population.

\subsection*{Accumulation of temporal variation}
For the simple birth process, contributions to the temporal variance [Eq.~(\ref{FPT_var})] fall off as the inverse square of the population size.
This inverse-square trend is also numerically observed in exactly inoculated age-structured population-growth models (Fig.~S3).
For populations with Poisson-distributed inocula the stochastic process of inoculation spontaneously generates temporal variation.
Thus, the largest contributions to temporal variation occur at small population sizes, which means that the growth rate at small population sizes {\color{black} should be} made manifest in the noise.

Changing perspective from small population sizes to early times, we next quantify the time scale over which temporal variance accumulates in a growing population.
We consider a two-step growth process. 
First, a population with inoculum size $n_0$ grows until a time $t$ according to the simple birth process, yielding a distribution $P_t(n \, | \, n_0)$ over abundances $N(t)$.
Second, at time $t$ population growth becomes deterministic and exponential (and hence this stage of growth does not contribute to the temporal variance).
We define the random variable $T[N(t)]$ to be the first-passage time for such deterministic exponential growth to reach a threshold $\Omega$ given that the inoculum size is a random variable $N(t)$,
\begin{equation}
    T[N(t)] = \frac{1}{\mu} \ln\left[\Omega/N(t) \right],
    \label{t_invert}
\end{equation}
where we assume the threshold $\Omega$ is much larger than any abundance $N(t)$ before deterministic growth begins.

The mean $\langle N(t) \rangle$ and variance $\langle N(t)^2 \rangle - \langle N(t) \rangle^2$ of the simple birth process are known [Eqs.~(\ref{mean_abundance}) and (\ref{P_var})], so the variance of this first-passage-time distribution may be computed with the delta method (\ref{delta_2}), yielding
\begin{equation}
\langle T[N(t)]^2 \rangle - \langle T[N(t)] \rangle^2 =
\frac{1}{\mu^2 n_0}\left(1 - e^{-\mu t} \right) + O\left(\frac{1}{n_0^2}\right). \label{answer}
\end{equation}
For $t \gg 1/\mu$, this recovers to leading order the relationship \eqref{temporal_variation_b} for the simple birth process between temporal standard deviation and inoculum size.
Strikingly, comparing Eq.~(\ref{answer}) to Eq.~(\ref{FPT_var}) (which was derived for growth that exclusively obeys the simple birth process), after a single division time $\ln(2)/\mu$ the temporal variance reaches half of its asympotic value.
{\color{black} Temporal variation is rapidly accumulated at early times (while populations are still small).}

\subsection*{Growth-rate inference}
Rearranging Eq.~(\ref{temporal_variation}), for a given inoculum size $n_0$ and experimentally measured TSD $\sigma_t$ at large threshold $\Omega$, either there are no other sources of noise and the growth rate is 
\begin{align}
   \mu_{\text{LB}} &= \frac{1}{\sigma_t}\left[ \frac{1}{n_0^2} + \frac{1}{(n_0+1)^2} + \cdots + \frac{1}{(\Omega-1)^2}\right]^{1/2}, \label{mu_LB}
\end{align}
or there are other sources of noise and the growth rate exceeds $\mu_{\text{LB}}$.
In general then, $\mu_{\text{LB}}$ is a lower bound for the growth rate,
{\color{black} so long as measurements are taken before abundance trajectories focus and decrease the noise ({\em e.g.,}  when they approach carrying capacity,  as in Fig.~\ref{growth_trajs}ab).}

\begin{figure}[t] 
    \includegraphics[width=0.48\textwidth]{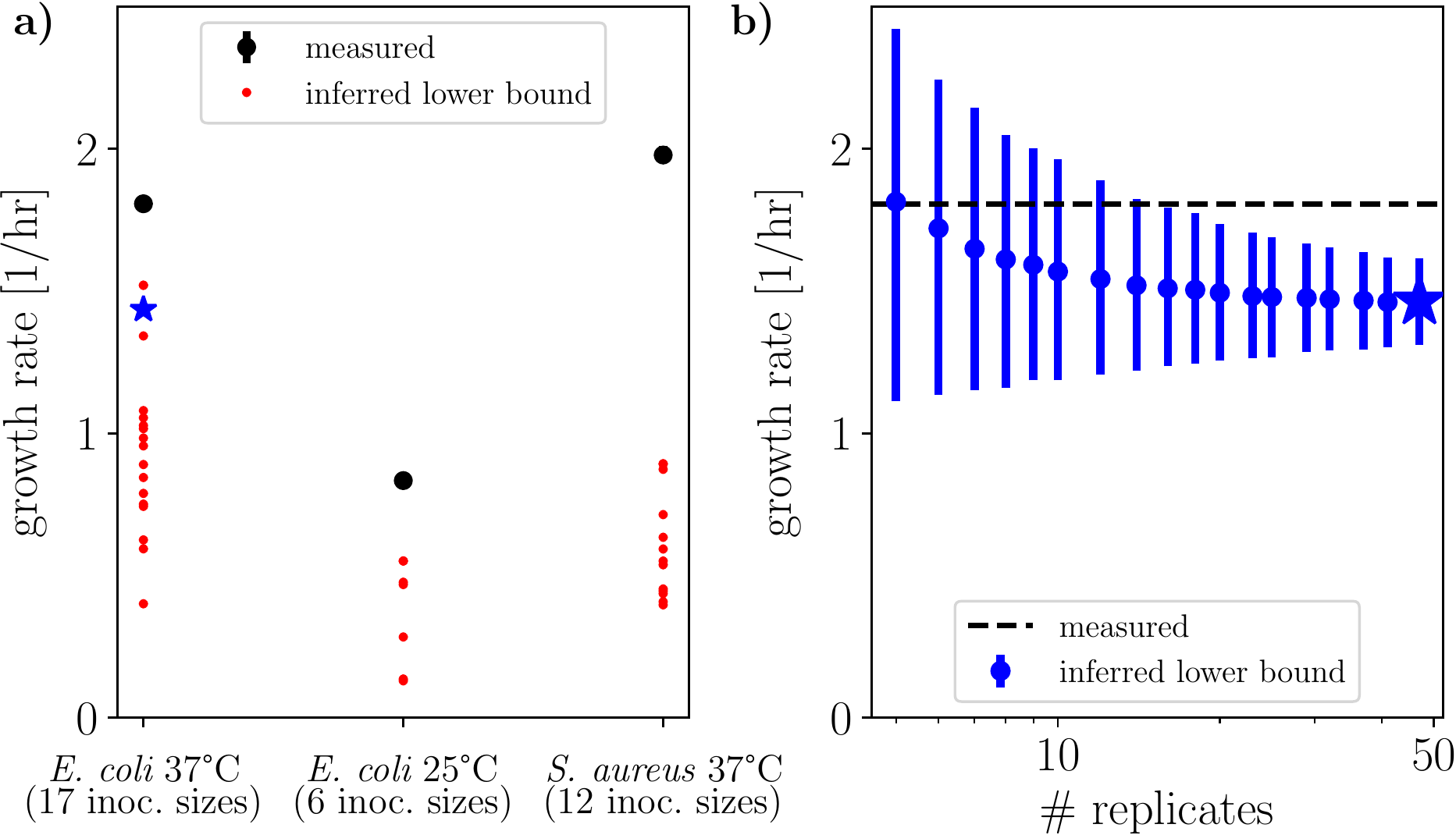}
    \caption{\textbf{Noise-based inference of growth-rate lower bounds.} 
    (a) Growth-rate lower bounds $\mu_{\text{LB}}$, inferred for each organism, growth condition, and inoculum size, are plotted as red dots.
    Measured growth rates (black points, as in Fig.~\ref{growth_trajs}c) are calculated as the slopes of log-transformed abundance trajectories.
    (b) Precision of growth-rate lower-bound inference, calculated by bootstrapping the abundance trajectories for {\em E. coli} at 37\textdegree C with mean inoculum size 2.8 [blue star in (a)].
    Error bars 
    for measured growth rates (a) and inferred lower bounds (b)
    show 68\% confidence intervals from $n$=5,000 bootstrap resamples per data point.
    \label{growth_inference}}
\end{figure}

Figure~\ref{growth_inference}a compares inferred growth-rate lower bounds $\mu_{\text{LB}}$ for each organism, growth condition, and inoculum size to the measured growth rate of each organism and growth condition.
The measured rate exceeded the greatest of the lower bounds by 19\% in {\em E.~coli} at 37\textdegree C, 51\% in {\em E.~coli} at 25\textdegree C, and 71\% in {\em S.~aureus} at 37\textdegree C.
To probe how confidence in the estimation of $\mu_{\text{LB}}$ depends on the number of replicate growth trajectories, we bootstrap resampled a set of 47 abundance trajectories with mean inoculum size 2.8 in Fig.~\ref{growth_inference}b.

Since most noise accumulates at small population sizes, the inferred growth-rate lower bound should be dominated by the growth rate at small population sizes.
This meets an important need in microbial ecology experiments, which is to measure the growth rate of strains before they significantly change the media.
Contemporary approaches quantify growth rates in small bacterial populations by directly observing the spatiotemporal dynamics of bacteria at sub-100nm spatial resolution, requiring cutting-edge microscopy and analysis methods \cite{Hartmann2019, Welker2021}.
By comparison, $\mu_{\text{LB}}$ depends exclusively on quantities that are straightforward to measure with standard microbiology lab equipment (namely, microplate readers and materials for colony-forming-unit counting assays).
{\color{black} Future bacterial growth experiments with dynamic growth rates could test this method's capacity to infer past growth rates from the noise at later times.}

\subsection*{Desynchronization of division times}
Finally we sought to understand when age-structured population growth becomes indistinguishable from the simple birth process, a crossover that helps to explain why TSDs of the two models have the same scaling behavior for large inoculum sizes in Fig.~\ref{fig:power_law}.
This crossover occurs when growth-rate oscillations in the age-structured model (corresponding to initially synchronized division events) desynchronize, at which point the population grows at a constant exponential rate \cite{Nisbet1986}.

In Supplementary Information Section C, we consider a deterministic age-structured population-growth model and apply Laplace-transform methods to determine the decay rate of growth-rate oscillations.
For a division-time distribution with 25-minute mean and 22\% coefficient of variation, the growth dynamics of a single inoculum asymptote to pure exponential growth after $\sim$3 division cycles (Fig.~S1).
Our bacterial optical-density measurements have a resolution of 0.001 ($\sim$$3 \times 10^5$ CFUs, corresponding to {\color{black} $\sim$18 division cycles)}, which suggests that such measurements cannot resolve any abundance oscillations predicted by age-structured growth models.
{\color{black} Said another way, after a few division cycles one may approximate the growth dynamics of age-structured growth by a simple birth process.}

We note that the deterministic age-structured model we consider ignores correlations between mother and daughter generation times, which have been empirically observed in bacteria \cite{LinAmir2017}.
Models that include cell-size control can extend the predicted persistence time of growth-rate oscillations \cite{Jafarpour2019}.
{\color{black} In the future, time-lapse microscopy of entire bacterial populations could be used to directly observe the desynchronization of populations with small inoculum sizes. }

\section*{Discussion}
Stochastic population growth, by its nature, produces a distribution of abundance trajectories over time \cite{Nisbet2003}.
For exponentially growing populations, the mean trajectory of this distribution contains information about the population growth rate, given by the slope of the log-transformed trajectory.
We demonstrated in this paper that the temporal standard deviation is a second statistic that reports on the population growth rate.
Temporal variation is especially informative when the birth rate is much larger than the death rate; {\color{black} temporal variation is less meaningful when populations fluctuate about a steady-state abundance or go extinct \cite{RichterDyn1972}.}

Traditionally it has been difficult to measure the growth rate of bacteria at small population sizes without expensive microscopy equipment, since conventional optical-density measurements are unable to resolve growth at small scales \cite{Hell2004, Hartmann2019, Bar2020, Welker2021}.
{\color{black} Addressing this need, our noise-based inference method suggests that the temporal standard deviation at a large population threshold (easily calculated with optical-density measurements) can be related to the growth rate at small population sizes.}


{\color{black} The difference between the directly measured growth rate and the noise-inferred growth rate is proportional to the extrinsic noise that is not accounted for by the stochastic growth model.
Candidate extrinsic noise sources include variability in media conditions, the duration of lag phase, and environmental conditions.
Future experiments that deliberately vary the strength of a single noise source could isolate that noise source's contribution to temporal variation.}

Temporal variation is a natural and useful description of noisy population growth.
{\color{black} It does not depend on the conversion factor between optical density and CFUs, saving experimental effort. }
{\color{black} The population dynamics of colonizing species during microbiome assembly are stochastic \cite{JonesLudington2022} and could be characterized in terms of temporal variation. 
}
The lower bound for the noise in {\em S.\ aureus} growth suggests a lower bound on the variation in times at which patients develop symptoms from the virulent hospital pathogen methycillin-resistant {\em S.\ aureus} (MRSA) following exposure \cite{VardakasFalagas2009}.

In an era of high-throughput biological experiments, noise-based analyses are becoming increasingly valuable. 
In this paper we found a signal in the noise that relates growth rate, inoculum size, and temporal standard deviation in exponentially growing systems.
{\color{black} Leveraging this relationship, in well-controlled bacterial growth experiments we demonstrated a proof of concept for the noise-based inference of population growth rate, setting the stage for future statistical analyses of noisy population growth.}

\matmethods{
\vspace{-2em}
\subsection*{Bacterial growth experiments} Either {\em E.~coli} strain MG1655 or {\em S.~aureus} strain NCTC 8532 was grown overnight in lysogeny broth (LB), then back diluted 1:1000 and grown to a 600nm optical density (OD600) of 0.5.
At this optical density bacterial growth is in mid-log phase.
Serial dilutions were performed to obtain a culture with cell concentrations between 1 and 150 CFU per 2$\mu$L.
This cell culture was subsequently used to inoculate bacterial growth experiments ({\em e.g.,} those in Fig.~\ref{growth_trajs}a,b) by pipetting 2$\mu$L of cell culture into 198 $\mu$L of LB media.
Pipetting was performed with the Rainin Pipet-Lite Multi Pipette L8-20XLS+, accurate to $\pm 0.2 \mu$L.
For each cell-culture concentration, 42 replicates were inoculated on the same 96-well plate to reduce variation, with 6 wells left as blank controls; each 96-well plate was inoculated with two sets of bacterial growth experiments. 
Plates were sealed with a ``breathe-easy'' with small holes poked in it to increase oxygen.
Preparation and inoculation of 96-well plates was performed at 24.6\textdegree C (room temperature).
Preparing each batch of experiments (consisting of three 96-well plates) took $\sim$15 minutes from start to finish, with inoculations for each inoculum size spanning $\sim$3 minutes from start to finish.

Plates were grown in a Biotek Epoch 2 plate reader for 24 hours at 37\textdegree C (or 25\textdegree C) with continuous orbital shaking.
Optical-density readings at OD600 were taken every two or three minutes.
When {\em E.~coli} was grown at 25\textdegree C, the time in the plate reader was extended to 48 hours. 
By the Beer-Lambert law, bacterial population size and OD600 are linearly correlated in the sensitivity range of the plate reader ($>$0.01 OD) \cite{Mira2022}.
Optical-density measurements therefore serve as a proxy for bacterial population size.

\subsection*{Measurement of inoculum size} For each concentration of cell culture, the distribution of the number of bacteria pipetted into each well of the 96-well plate ({\em i.e.,} the inoculum size) was inferred by spot plating identical volumes of cell culture on LB-agar plates \cite{Gaudy1963}.
Colonies were counted after 16 hours of growth.
For each concentration of cell culture, the inoculum size is roughly Poisson-distributed (Fig.~S2).
The mean $n_0$ of nonzero inoculum sizes is utilized in Figs.~3 and 4.

{\color{black}
\subsection*{Lag phase}
For the three inoculum sizes in Fig.~\ref{growth_trajs}a we do not find evidence of a significant lag phase: the calculated time for a model of deterministic exponential growth with no lag phase to reach an optical density of 0.03 ($\sim$$1.4\times10^7$ CFUs) exceeded the mean observed time by 30 min for $n_0 = 80.7$; by 36 min for $n_0 = 16.1$; and by 59 min for $n_0 = 1.8$.

This analysis required a standard curve to convert optical density measurements to CFUs, measured by spot plating following serial dilution  \cite{StevensonPilizota2016}.
For this standard curve, measured optical densities spanned from 0.01 to 0.6, and measured CFUs spanned from $6 \times 10^6$ to $2 \times 10^8$.
For each cell-culture concentration, measurements were performed for 7 biological replicates.
Based on linear regression, an OD of 0.03 corresponds to $\sim$$1.4\times 10^7$ CFUs.
}

\subsection*{Bacterial strains} The MG1655 strain of {\em E.~coli} (ATCC 700926) was obtained from the Broderick lab at Johns Hopkins University.
The NCTC 8532 strain of {\em S.~aureus} (ATCC 12600) was obtained from the Saleh lab at Johns Hopkins University.
Cultures were obtained by streaking from glycerol stocks onto LB-agar plates and grown for 16 hours at 37\textdegree C.

\subsection*{Criteria for omission of growth curves} 
Bacterial growth curves were omitted from analysis if: (i) a well was missing an air puncture, causing anerobic growth (3/1439 replicates omitted), (ii) a well was contaminated (2/1439 replicates omitted), or (iii) raw OD600 after 1 hour of growth was above 0.125, indicating initial condensation or measurement error (47/1439 replicates omitted).
In total, these exclusion criteria led to the omission of 4\% (52/1439) of growth trajectories.
Figure~S4 shows all raw growth curves, with omitted curves in red.

\subsection*{Removing optical-density background} The measurement background---corresponding to the light occluded by solution (not bacteria) in a well---was subtracted from each optical-density time-series.
The background was calculated as the mean optical density at time 0 for each 96-well plate, and ranged from an optical density of 0.099 to 0.121.
Figures~\ref{growth_trajs}a and \ref{growth_trajs}b show representative background-subtracted optical-density measurements.
For reference, empty dry wells yield optical-density measurements of 0.005.

\subsection*{Growth-rate calculation} For a particular bacterial growth curve, the growth rate $\mu$ is determined by linearly regressing the log-transformed background-subtracted optical-density trajectory.
Operationally, the growth rate at a given time $t_0$ is calculated as the slope of the best-fit line for the 30-minute window centered at $t_0$.
A single growth rate was calculated for each organism and growth condition, defined as the average growth rate across replicates and inoculum sizes evaluated at times $t_0$ when optical-density trajectories reach threshold optical density 0.03: {\em E.~coli} at 37\textdegree C grows at $\mu=1.8$/hr, {\em E.~coli} at 25\textdegree C grows at $\mu=0.8$/hr, and {\em S.~aureus} at 37\textdegree C grows at $\mu=2.0$/hr.
The growth rate is relevant for plotting TSDs in units of division time in Fig.~\ref{fig:power_law}, since an organism's division time is defined as $\ln(2)/\mu$.

\subsection*{Population-growth models} For each population-growth model plotted in Fig.~\ref{model_TSDs}, a set of integer inoculum sizes ranging from 1 to 30 were simulated.
Models with Poisson-distributed inocula used this integer inoculum size as the Poisson shape parameter; the subsequent zero-truncated Poisson distribution has a larger mean inoculum size, giving rise to non-integer mean inoculum sizes.
The simple birth process with exact inoculation (red) and deterministic exponential growth with Poisson-distributed inocula (blue) were computed exactly with Eqs.~(10a) and (S26), respectively.

The age-structured population-growth model with exact inoculation (gold) was simulated in an agent-based manner.
Inoculated individuals were assumed to be at a random point along their division cycle, so their first division event was set to a random time uniformly drawn from [0, $(\ln 2)/\mu$].
Thereafter, after each division event, the two resulting individuals each randomly drew their next division time from a division-time distribution that is determined by a 20-stage growth process (in which reaching the next stage of development is a Poisson process with constant rate): specifically, this growth process yields a division-time distribution given by a chi-squared distribution $\chi^2(40)$ \cite{Kendall1948}, linearly rescaled so the mean division time was 25 minutes.
Simulated TSDs were calculated at a threshold of 500 individuals.

Lastly, simple-birth-process simulations with Poisson-distributed inocula (purple) were performed by drawing 2,000 inoculum sizes from an appropriate Poisson distribution, then performing stochastic simulations using the Python function \texttt{birdepy.simulate.discrete}.
For each set of simulations (gold, green, purple), 95\% confidence intervals were computed by bootstrapping using the Python function \texttt{scipy.stats.bootstrap}. 

\subsection*{Deterministic model of age-structured growth} Simulations of the deterministic age-structured population-growth model displayed in Fig.~S1 were performed using the Mathematica functions \texttt{TransferFunctionModel}, \texttt{TransferFunctionPoles}, and \texttt{NInverseLaplaceTransform}.

\subsection*{Data and software availability} Raw data from bacterial growth experiments and software that can recreate main text figures are available online at GitHub: \texttt{https://github.com/erijones/intrinsic\_variation}.
Analyses were performed with Python (version 3.9.7) and Mathematica (version 12.1.0.0).
}
\showmatmethods{}

\acknow{We thank Robert Scheffler and Ferdinand Pfab for helpful discussions.
Support was provided by Banting and Pacific Institute for the Mathematical Sciences Postdoctoral Fellowships (E.W.J.); National Science Foundation IOS 2032985 (W.L.); Natural Sciences and Engineering Research Council of Canada (NSERC) Discovery Grant RGPIN-2020-04950 and a Tier-II Canada Research Chair CRC-2020-00098 (D.A.S.); and the Carnegie-Canada Foundation (W.L.\ and D.A.S.).}
\showacknow{}

\newpage
\onecolumn

\begin{center}
    \huge{Supplementary Information} \\ 
\end{center}


\renewcommand{\theequation}{S\arabic{equation}}
\renewcommand{\thefigure}{S\arabic{figure}}
\setcounter{figure}{0}
\setcounter{equation}{0}

\section*{Section A: First-passage-time distribution of the simple birth process}
For a simple birth process with with probability $P_t(n \, | \, n_0)$ of a population consisting of $n$ individuals at time $t$ given an inoculum size of $n_0$,
the reaction probability $R_{\Omega}(t \, | \, n_0)$ that at time $t$ the population
size is greater than or equal to population threshold $\Omega$ is 
\begin{align}
    R_{\Omega}(t \, | \, n_0) &= 1 - \sum_{i=n_0}^{\Omega-1} P_{t}(i \, | \, n_0).
\end{align}
Since abundance trajectories are monotonic, the reaction probability is also related to the
first-passage-time probability $P_{\Omega}^{\text{FP}}(t \, | \, n_0)$ of times $t$ at which an abundance trajectory first reaches $\Omega$ individuals,
\begin{align}
    R_\Omega(t \, | \, n_0) &= \int_0^t P_{\Omega}^{\text{FP}}(\tau \, | \, n_0)
    \, \text{d}\tau.
\end{align}
By the fundamental theorem of calculus, the first-passage-time distribution $P_{\Omega}^{\text{FP}}(t \, | \, n_0)$
is related to the solution 
$P_t(n \, | \, n_0)$ of the simple birth process:
\begin{subequations}
\begin{align}
    P_n^{\text{FP}}(t \, | \, n_0) &= - \sum_{i=n_0}^{n-1} \frac{\text{d} P_t(i
    \, | \, n_0)}{\text{d}t} \label{prefinal_answer} \\
    &= 
    \mu(n-n_0) {n-1 \choose n_0 - 1}
    e^{-\mu n_0 t}(1 - e^{-\mu t})^{n-n_0-1} . \label{final_answer}
\end{align}
\end{subequations}

\vspace{4em} 
Next, we prove that Eqs.~(\ref{prefinal_answer}) and (\ref{final_answer}) are equal. Recall that 
\begin{align}
P_{t}(n \, | \, n_0) &= \frac{(n-1)!}{(n_0 - 1)! (n-n_0)!} e^{-\mu n_0 t} (1 -
    e^{-\mu t})^{n-n_0},
\end{align}
so by Eq.~(\ref{prefinal_answer}),
\begin{subequations}
\begin{align}
    P_{n}^{\text{FP}}(t \, | \, n_0) &=
    \frac{\text{d}}{\text{d}t} \left[ -e^{-\mu n_0 t} (1 - e^{-\mu t})^{-n_0} \sum_{i=n_0}^{n-1} {i-1 \choose n_0-1}(1 -
    e^{-\mu t})^i\right]  \\
    &= \mu e^{-\mu n_0 t} (1 - e^{-\mu t})^{-n_0 - 1} \nonumber\\
    &\quad \times  
    \left[ \sum_{i=n_0}^{n-1} {i-1 \choose n_0 - 1} (n_0 - i e^{-\mu t}) (1 - e^{-\mu t})^i 
    \right] \\
    &= \mu e^{-\mu n_0 t} (1 - e^{-\mu t})^{-n_0 - 1} \nonumber\\
    &\quad \times  
    \frac{1}{(n_0 - 1)!}\left[ \sum_{i=n_0}^{n-1} \frac{(i-1)!}{(i-n_0)!}(n_0 - i e^{-\mu t}) (1 - e^{-\mu t})^i 
    \right].
\end{align}
\end{subequations}
%
To evaluate the quantity in square brackets, we proceed by induction. Define
\begin{align}
    a_i &\equiv {i-1 \choose n_0 - 1} (n_0 - i e^{-\mu t}) (1 - e^{-\mu t})^i.
\end{align}
We will show that \begin{align}S_{n_0,n} \equiv \sum_{i=n_0}^{n-1} a_i = {n-1 \choose n_0 - 1}
(1 - e^{-\mu t})^n (n-n_0).\end{align}
First, the base case is satisfied:
\begin{align}
    S_{n_0,n_0+1} = a_{n_0} = n_0(1 - e^{-\mu t})^{n_0+1}.
\end{align}
Next we assume 
\begin{align}
    S_{n_0, n-1} = {n-2 \choose n_0 - 1}
    (1 - e^{-\mu t})^{n-1} (n-1-n_0)
\end{align}
and prove the inductive step 
\begin{align}
S_{n_0,n} = {n-1 \choose n_0 - 1}
(1 - e^{-\mu t})^n (n-n_0).
\end{align}
We find
\begin{subequations}
\begin{align}
    S_{n_0,n} &= S_{n_0,n-1} + a_{n-1} \\
    &= {n-2 \choose n_0 - 1}
    (1 - e^{-\mu t})^{n-1} (n-1-n_0) + {n-2 \choose n_0-1}(n_0 - (n-1)e^{-\mu t})(1
    - e^{-\mu t})^{n_0-1} \\
    &= {n-2 \choose n_0 - 1}
    (1 - e^{-\mu t})^{n-1} \left[
        (n-1-n_0) + (n_0 - (n-1)e^{-\mu t}) \right]\\
    &= {n-2 \choose n_0 - 1}
    (1 - e^{-\mu t})^{n-1} \left[
        n-1- n e^{-\mu t} + e^{-\mu t} \right]\\
    &= {n-2 \choose n_0 - 1}
    (1 - e^{-\mu t})^{n} (n-1) 
        \\
    &= {n-1 \choose n_0-1}
    (1 - e^{-\mu t})^{n} (n-n_0),
\end{align}
\end{subequations}
as required. Therefore,
\begin{subequations}
\begin{align}
    P_{n}^{\text{FP}}(t \, | \, n_0) &= \mu e^{-\mu n_0 t}(1 - e^{-\mu
    t})^{-n_0-1} S_{n_0,n} \\
    &= 
    \mu(n-n_0) {n-1 \choose n_0 - 1}
    e^{-\mu n_0 t}(1 - e^{-\mu t})^{n-n_0-1} ,
\end{align}
\end{subequations}
in agreement with Eq.~(\ref{final_answer}).

\newpage

\section*{Section B: Mean and variance of the first-passage-time distribution for the simple birth process}
The mean first-passage time is
\begin{subequations}
\begin{align}
    \langle t \rangle_{n \, | \, n_0} &= \int_0^{\infty} t\,  P_n^{\text{FP}}(t \,
    | \, n_0) \, \text{d}t \\
    &= \mu(n-n_0) {n-1 \choose n_0 - 1} \int_0^\infty t e^{-\mu n_0 t}(1 -
    e^{-\mu t})^{n-n_0-1}  \, \text{d}t \\
    &= \mu(n-n_0) {n-1 \choose n_0 - 1} \sum_{k=0}^{n-n_0-1} {n-n_0-1 \choose
    k} (-1)^k \int_0^\infty t
    e^{-\mu (n_0 + k) t} \, \text{d}t \\
    &= \mu(n-n_0) {n-1 \choose n_0 - 1} \sum_{k=0}^{n-n_0-1}  {n-n_0-1 \choose
    k}(-1)^k 
    \left[\frac{- e^{-\mu (n_0 + k)t} (1 + \mu(n_0 +
    k)t)}{\mu^2(n_0+k)^2}\right]_0^{\infty} \\
    &= \mu(n-n_0) {n-1 \choose n_0 - 1} \sum_{k=0}^{n-n_0-1} 
    {n-n_0-1 \choose
    k}\frac{(-1)^k}{\mu^2(n_0+k)^2} \\
    &= \frac{1}{\mu} \left(\frac{1}{n_0} +
    \frac{1}{n_0+1} + \cdots + \frac{1}{n-1}\right), \label{mfpt}
\end{align}
\end{subequations}
where the last equality follows from the identity Eq.~(\ref{binomial_identity}).
Similarly,
\begin{subequations}
\begin{align}
    \langle t^2 \rangle_{n \, | \, n_0} &= \int_0^{\infty} t^2 \,  P_n^{\text{FP}}(t \,
    | \, n_0) \, \text{d}t \\
    &= \mu(n-n_0) {n-1 \choose n_0 - 1} \sum_{k=0}^{n-n_0-1} {n-n_0-1 \choose
    k} (-1)^k \int_0^\infty t^2
    e^{-\mu (n_0 + k) t} \, \text{d}t \\
    &= \mu(n-n_0) {n-1 \choose n_0 - 1} \sum_{k=0}^{n-n_0-1}  {n-n_0-1 \choose
    k}(-1)^k 
    \left[\frac{- e^{-\mu (n_0 + k)t} (2 + 2 \mu(n_0 +
    k)t + \mu^2 (n_0+k)^2 t^2)}{\mu^3(n_0+k)^3}\right]_0^{\infty} \\
    &= 2 \mu(n-n_0) {n-1 \choose n_0 - 1} \sum_{k=0}^{n-n_0-1} 
    {n-n_0-1 \choose
    k}\frac{(-1)^k}{\mu^3(n_0+k)^3} \\
    &= \frac{1}{\mu^2}\left[
    \left(\frac{1}{n_0} +
    \frac{1}{n_0+1} + \cdots + \frac{1}{n-1}\right)^2
    + \left( \frac{1}{n_0^2} +
    \frac{1}{(n_0+1)^2} + \cdots + \frac{1}{(n-1)^2} \right) \right], \label{t2}
\end{align}
\end{subequations}
where the last equality follows from the identity Eq.~(\ref{id2}). Thus, the temporal variance
$\sigma^2_t \equiv \langle t^2 \rangle - \langle t \rangle^2$ is
\begin{align}
    \sigma^2_t = 
    \frac{1}{\mu^2} \left( \frac{1}{n_0^2} +
    \frac{1}{(n_0+1)^2} + \cdots + \frac{1}{(n-1)^2} \right).
\end{align}

\vspace{4em}
To derive the identities Eqs.(\ref{binomial_identity}) and (\ref{id2}), start from the identify \cite{Knuth1997}
\begin{equation}
\sum_{k=0}^n \left( \begin{matrix} n \\ k \end{matrix} \right) \frac{(-1)^k}{k + x} =
\left[x \left( \begin{matrix} n + x \\ n \end{matrix} \right) \right] ^{-1}.
\end{equation}
Differentiating  with respect to $x$ yields
\begin{subequations}
\begin{align}
\sum_{k=0}^n \left( \begin{matrix} n \\ k \end{matrix} \right) \frac{(-1)^k}{(k + x)^2} &=
-\frac{\text{d}}{\text{d}x} \left[\frac{n!}{x(x+1)\cdots(x+n)}\right] \\
&=\frac{n!}{x(x+1)\cdots(x+n)} \left(\frac{1}{x} + \frac{1}{x+1} + \cdots + \frac{1}{x+n}\right) . 
\label{binomial_identity}
\end{align}
\end{subequations}
Differentiating the identity Eq.~(\ref{binomial_identity}) again gives
\begin{subequations}
\begin{align}
2 \sum_{k=0}^n \left( \begin{matrix} n \\ k \end{matrix} \right) \frac{(-1)^k}{(k + x)^3} &=
\frac{\text{d}^2}{\text{d}x^2} \left[\frac{n!}{x(x+1)\cdots(x+n)}\right] \\
&=\frac{n!}{x(x+1)\cdots(x+n)} \left(\frac{1}{x} + \frac{1}{x+1} + \cdots + \frac{1}{x+n}\right)^2 \\
&\quad +\frac{n!}{x\cdots(x+n)} \left(
\frac{1}{x^2} + \frac{1}{(x+1)^2} + \cdots + \frac{1}{(x+n)^2}
\right). \label{id2}
\end{align}
\end{subequations}
These two identities provide the simplifications needed for Eqs.~(\ref{mfpt}) and (\ref{t2})

\newpage

\section*{Section C: Deterministic age-structured population growth}

Organismal division is intricately choreographed and can often be broken down into discrete stages \cite{Keyfitz1997}.
Here we examine deterministic age-structured population growth models in which division-time distributions describe the timing of division events.
In particular, with Laplace-transform methods we characterize the desynchronization of initially synchronized division events.

Let $n(a,t)\, \text{d}a$ be the number of individuals aged between $a$ and $a + \text{d}a$ at time $t$ (where age is defined as elapsed time since previous division), and assume individuals divide with propensity $\beta(a)$.
Population dynamics are governed by the PDE \cite{Nisbet1986, Keyfitz1997}
\begin{equation} 
    \frac{\partial n}{\partial t}+\frac{\partial n}{\partial a}+\beta (a)n=0, \label{pde}
\end{equation}
together with the renewal condition that describes how individuals divide,
\begin{equation} \label{renew}
    R(t)\equiv n(0,t)=2\int\limits_{0}^{\infty }{\beta (a)}n(a,t)\text{d}a,
\end{equation}					
and the initial condition 
\begin{equation}
 n(a,0)={{n}_{0}}(a),  \label{ic}
\end{equation}
where $R(t)$ is the recruitment rate of newly divided individuals at time $t$ ({\em i.e.,} the rate at which age 0 individuals enter the population, or roughly twice the growth rate of the simple birth process).

A formal solution to Eq.~(\ref{pde}) is
\begin{equation} \label{formal}
n(a,t)=\begin{cases} 
 R(t-a)S(a) \text{         for } t>a  \\
   {{n}_{0}}(a-t){\tilde{S}(a, t) } \text{   for }t\le a\text{            }  \\
\end{cases} ,
\end{equation}
where the ``survival'' function $S(a) \equiv \exp \left[ -\int_{0}^{a}{\beta (u) \, \text{d}u} \right]$ is the proportion of individuals that survive to age $a$ before dividing, and the modified survival function $\tilde{S}(a, t) \equiv \exp \left[ -\int_{a-t}^{a}{\beta (u) \, \text{d}u} \right]$
is the proportion of individuals that survive to age $a$ before dividing given that they existed and were undivided at age $a-t$.
It is also convenient to define a normalized division-time distribution $\text{P}_{\text{DT}} (a) \equiv \beta (a)S(a)$.  
Provided no deaths occur, these functions are related according to
\begin{equation}
    S(a)=1-\int_{0}^{a}{\text{P}_{\text{DT}} (u)\, \text{d}u}.
\end{equation}

\begin{figure}[t] 
    \includegraphics[width=\textwidth]{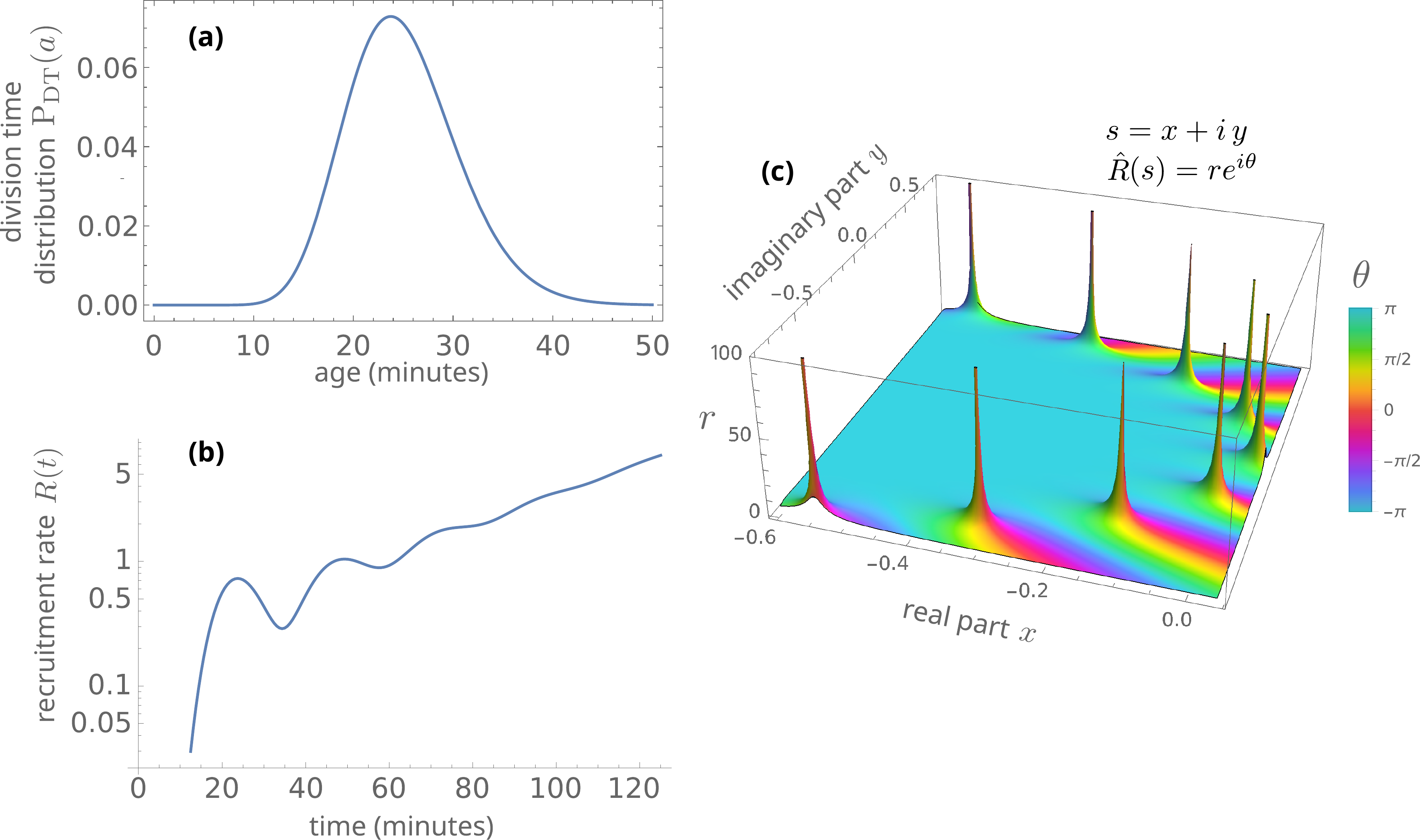}
    \caption{
    \textbf{Characterization of age-structured population growth.} 
    (a) Division-time distribution for a 20-stage population-growth model given by a chi-squared distribution $\chi^2(2k)$ with $k=20$ (22\% coefficient of variation), linearly rescaled such that the mean division time is 25 minutes.
    (b) Recruitment rate $R(t)$ contains transient oscillations that decay after a few division times. 
    (c) Poles of the Laplace transformed recruitment rate $\hat{R}(s)$ in an age-structured population-growth model, plotted in the complex plane.
    The pole with the largest real part has a positive real part and determines the long-run population-growth rate; the complex-conjugate pairs contribute to transient oscillations.
    The location of the three poles nearest the right-hand edge are used to compute the coherence number Eq.~(\ref{coherence}).
\label{PSPM_plots}}
\end{figure}

From Eqs.~(\ref{renew}) and (\ref{formal}),
\begin{align} \label{Lotka}
 R(t)&=\underbrace{2\int_{0}^{t}{R(t-a)\beta (a)S(a)\, \text{d}a}}_{\text{created after }t=0\text{ }}+2\underbrace{\int_{t}^{\infty }{{{n}_{0}}(a-t)\beta (a)\tilde{S}({a, t})\, \text{d}a}}_{\text{from inoculum}} \nonumber \\
  &=2\int_{0}^{t}{R(t-a)\text{P}_{\text{DT}} (a)\, \text{d}a}+F(t),
\end{align}
with $F(t) \equiv 2\int_{t}^{\infty }{{{n}_{0}}(a-t) \beta(a) \tilde{S}(a, t) \, \text{d}a}$ giving contributions to the recruitment rate from individuals that have not divided since inoculation.

In the special case where the inoculum consists of $N_0$ newly divided cells, ${{n}_{0}}(a)={N_0 {\delta}}(a)$ for Dirac delta function $\delta(a)$, and  $F(t)=2{{N}_{0}}\text{P}_{\text{DT}} (t)$.
From Eqs.~(\ref{renew}) and (\ref{Lotka}), the total population size is
\begin{equation}
N(t)=\int_{0}^{\infty }{n(a,t)\, \text{d}a=\underbrace{\int_{0}^{t}{R(t-a)S(a)\, \text{d}a}}_{\text{created after }t=0}+\underbrace{\int_{t}^{\infty }{{{n}_{0}}(a-t)\tilde{S}({a, t})\, \text{d}a}}_{\text{from inoculum}}},
\end{equation}
which in this special case simplifies to
\begin{equation}
N(t)=\int_{0}^{t}{R(t-a)S(a)\, \text{d}a}+{{N}_{0}}S(t).
\end{equation}
 
The dynamics of this age-structured population therefore depend entirely on the division propensity $\beta(a)$, by way of the survival function $S(a)$ and the division-time distribution $\text{P}_{\text{DT}}(a)$. 
	
Laplace transforming Eq.~(\ref{Lotka}) yields
\begin{equation}
    \hat{R}(s)=2\hat{R}(s)\hat{\text{P}}_{\text{DT} }(s)+\hat{F}(s),
\end{equation}
implying that 
\begin{equation}
\hat{R}(s)=T(s)\hat{F}(s), \label{B}
\end{equation}
with $T(s) \equiv 1/[1-2\hat{\text{P}}_{\text{DT} }(s)]$.
The transfer function $T(s)$ describes the mapping in the complex $s$-plane from both the initial distribution of ages in the population and the division-time distribution to the solution of the dynamical system.
The explicit time series for a specified initial condition is obtained by inverse Laplace transformation, a task slightly simplified in the special case where the initial population consists of newly divided cells, for which $\hat{F}(s)\text{= }2{{N}_{0}}\hat{\text{P}}_{\text{DT} }(s)$.	

Following Kendall's 1948 seminal work \cite{Kendall1948}, we consider a class of age-structured models in which the division-time distribution for a $k$-stage population-growth model is given by a chi-squared distribution with $2k$ degrees of freedom, $\text{P}_{\text{DT}}(a) = \chi^2(2k)$.
The solution has the form
\begin{equation}
   R(t)=\sum\limits_{\text{all poles }i}{{{c}_{i}}\exp ({{s}_{i}}t)},
\end{equation}
where the coefficients $c_i$ depend on the initial age distribution, and the exponents $s_i$ are the locations in the complex $s$-plane of the poles of the transfer function. 
The pole $s_0$ with the largest real part 
determines
the long-run population-growth rate:
\begin{equation}
    R(t)={{c}_{0}}\exp ({{s}_{0}}t)\left( 1+\sum\limits_{\text{ }i\ne 0}{\frac{{{c}_{i}}}{{{c}_{0}}}\exp [({{s}_{i}}-{{s}_{0}})t]} \right), \label{sum}
\end{equation}
where the expression in parentheses approaches 1 as $t\to \infty $.
The subdominant poles $s_{1,2}$ are typically a complex-conjugate pair and characterize the approach to asymptotic exponential growth.
Defining ${{s}_{1,2}} \equiv  \sigma \pm i\omega $ for real $\sigma$, 
the leading terms in the summation in Eq.~(\ref{sum}) are proportional to $\exp [(\sigma -{{s}_{0}})t]\cos (\omega t-\phi )$, where $\phi$ sets the phase of any transient oscillations.  
The period of any transient oscillations is $2\pi /\omega $.  
For all cases we explored, this period is very close to the mean division time.

The transient decays by a factor of $e$ over a time interval $1/(s_0 - \sigma)$.
We define the {\em coherence number} $n_c$  as the number of oscillations before the transient decays by a factor of $e$,
\begin{equation}
     {{n}_{c}} \equiv \frac{\omega }{2\pi ({{s}_{0}}-\sigma )}.\label{coherence}
\end{equation}
The coherence number measures the rate at which intrinsic variability desynchronizes initially synchronized division events, and therefore informs the point at which detailed non-Markovian models may be approximated by coarse-grained Markovian models like the simple birth process.
Accordingly, it takes ${{\log }_{e}}10\times {{n}_{c}}\approx 2.3{{n}_{c}}$ oscillations in order for the transient to drop to 10\% of its original magnitude. 

Figure~\ref{PSPM_plots} illustrates these concepts for the ``deterministic skeleton'' \cite{Higgins1997} of the stochastic age-structured model used in Fig.~2, which describes the deterministic and incremental development of individuals until division.
It shows the division-time distribution for a 20-stage population-growth model with a mean division time of 25 minutes (Fig.~\ref{PSPM_plots}a), the growth rate over time (Fig.~\ref{PSPM_plots}b), and the leading poles ({\em i.e.,} those with largest real part) of the transfer function (and equivalently of $R(s)$) (Fig.~\ref{PSPM_plots}c).
The coherence number for this system is $n_c = 1.01$, implying that the approach to exponential growth (when the transients  have dropped to 10\% of their original magnitude) requires $\sim$2.3 cell division cycles. 

We conclude that the dynamics of the 20-stage model (with initially oscillatory growth rates) approach the dynamics of the simple birth process (with growth rates proportional to population size) after a few division cycles.

\newpage

\begin{figure}[t] 
    \includegraphics[width=.9\textwidth]{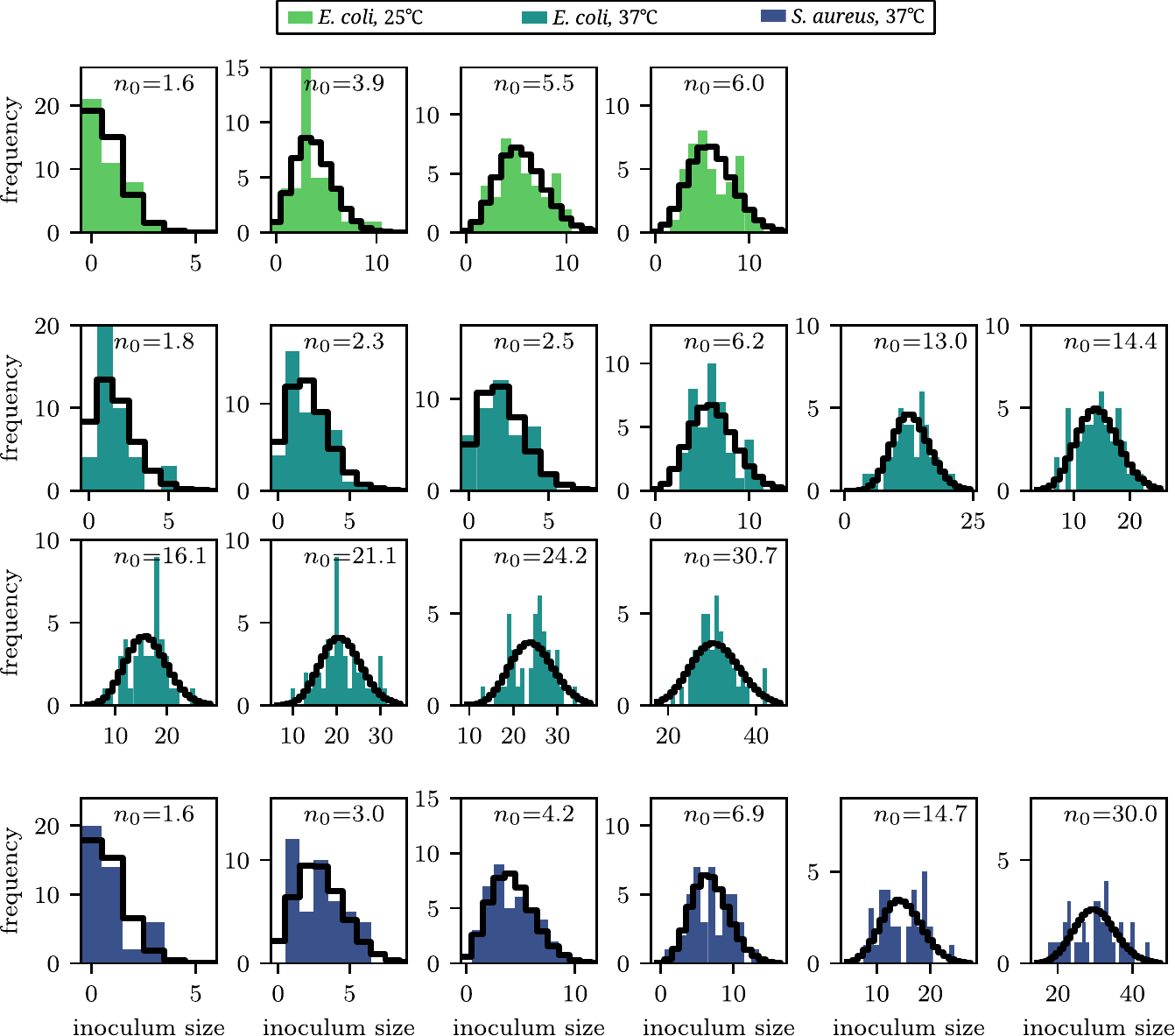}
    \caption{\textbf{Inoculum sizes are roughly Poisson-distributed.}
    Distribution of inoculum sizes (filled histogram), measured by spot plating, for 20 cell cultures of varying concentrations (Methods).
    Black: theoretical Poisson distribution for the measured mean inoculum size.
    Spot-plating experiments were performed for each organism and growth condition, as indicated by the legend.
    For each distribution we report the zero-truncated mean abundance $n_0$, plotted in Figure 4.
    \label{FigS1}}
\end{figure}

\phantom{text}
\newpage

\begin{figure}[t] 
    \includegraphics[width=\textwidth]{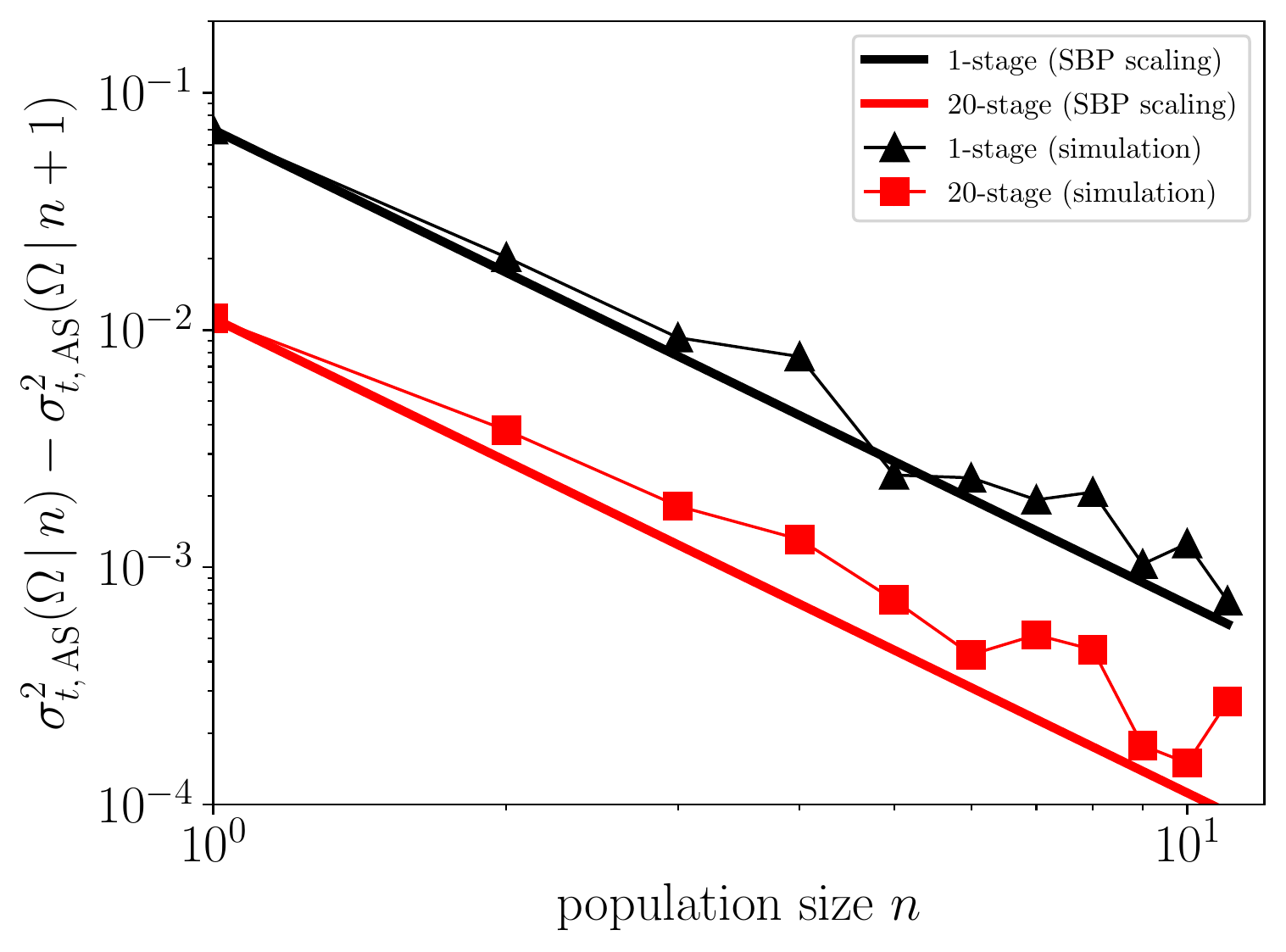}
    \caption{\textbf{Contributions to the temporal variance across population sizes for age-structured population models.}
    Difference in asymptotic (large-$\Omega$) temporal variance $\sigma^2_{t,\text{AS}}(\Omega \, | \, n_0)$ between inoculum sizes $n$ and $n+1$ ({\em i.e.}, the reduction in the asymptotic temporal variance by starting with one more individual) for age-structured population growth.
    Points are from 1-stage and 20-stage stochastic age-structured population models (Methods) with 20,000 trajectories.
    Temporal variances are evaluated at a threshold population size $\Omega = 500$.
    95\% confidence intervals are smaller than symbols.
    In the simple birth process, the temporal variance [Eq.~(10)] is a sum with summands that scale as $1/n^2$.
    Solid lines depict this $1/n^2$ scaling, starting from the contribution to the asymptotic temporal variance of a single individual (graphically, a $1/n^2$ power law starting from the $n=1$ data point).
    Thin lines are a guide to the eye.
    \label{FigS3}}
\end{figure}
\phantom{text}
\newpage

\begin{figure}[t] 
    \includegraphics[width=\textwidth]{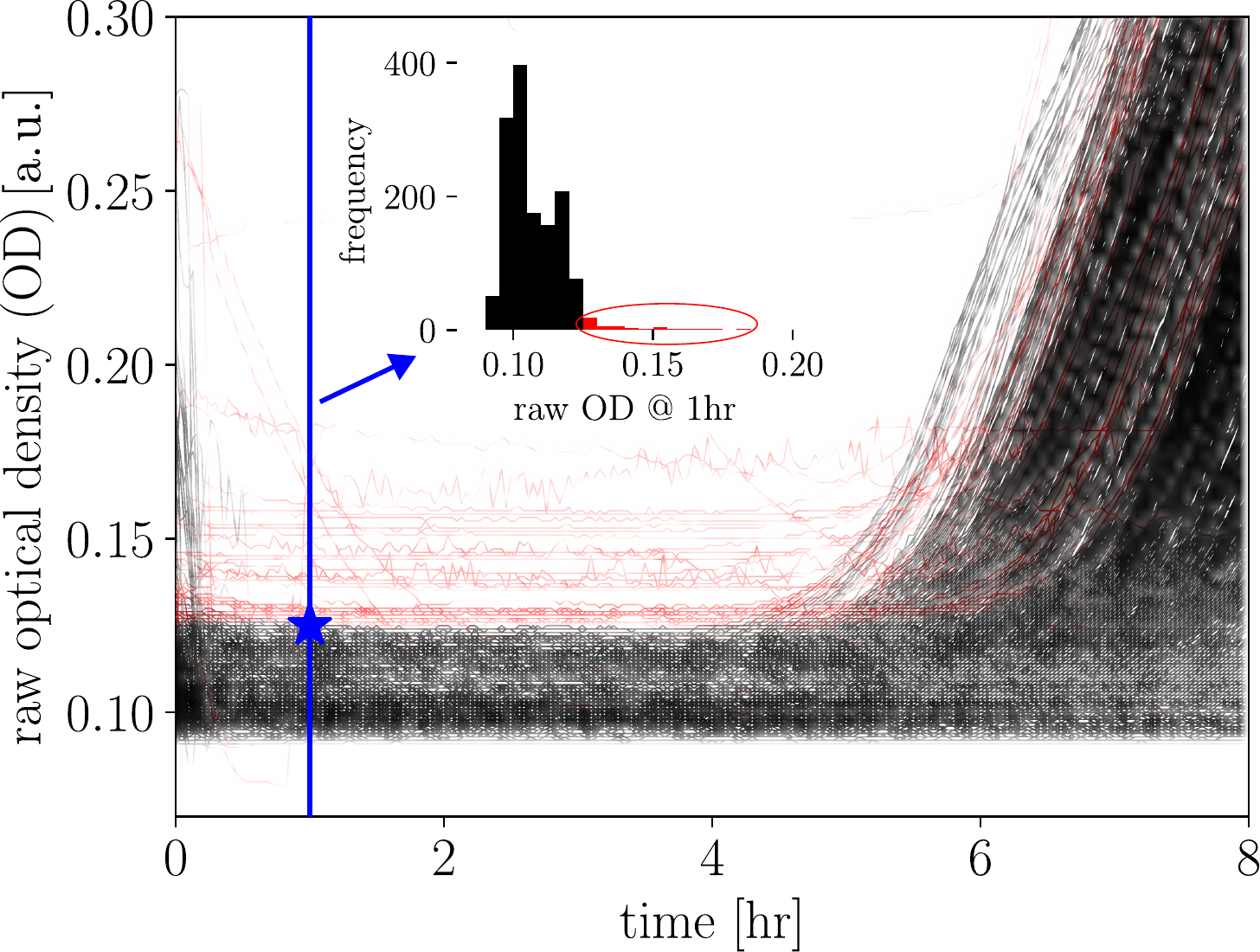}
    \caption{\textbf{Raw optical-density measurements from all 1434 bacterial growth experiments, with 47 excluded growth curves shown in red.}
    Growth curves with a raw optical density ({\em i.e.,} before the background has been subtracted) greater than 0.125 at 1 hour post inoculation (indicated by the blue star and line) were omitted from subsequent analysis, and are indicated in red.
    Including every growth curve marginally increases TSDs ({\em e.g.}, in Fig.~4), as outliers inflate trajectory spread.
    (inset) Histogram of raw optical density at 1 hour post inoculation; the circled red bins indicate the $\sim$4\% of trajectories that were excluded from subsequent analysis. 
    \label{FigS4}}
\end{figure}

\bibliographystyle{unsrt}

\begin{thebibliography}{10}

\bibitem{Hsieh2006}
Ch Hsieh, et~al., Fishing elevates variability in the abundance of exploited
  species.
\newblock {\em\protect\JournalTitle{Nature}} \textbf{443}, 859--862 (2006).

\bibitem{Stukalin2013}
EB Stukalin, I Aifuwa, JS Kim, D Wirtz, SX Sun, Age-dependent stochastic models
  for understanding population fluctuations in continuously cultured cells.
\newblock {\em\protect\JournalTitle{Journal of The Royal Society Interface}}
  \textbf{10}, 20130325 (2013).

\bibitem{Kendall1948}
DG Kendall, On the role of variable generation time in the development of a
  stochastic birth process.
\newblock {\em\protect\JournalTitle{Biometrika}} \textbf{35}, 316--330 (1948).

\bibitem{Powell1955}
EO Powell, Some features of the generation times of individual bacteria.
\newblock {\em\protect\JournalTitle{Biometrika}} \textbf{42}, 16--44 (1955).

\bibitem{Powell1958}
EO Powell, An outline of the pattern of bacterial generation times.
\newblock {\em\protect\JournalTitle{Microbiology}} \textbf{18}, 382--417
  (1958).

\bibitem{Waugh1974}
WAO Waugh, Asymptotic growth of a class of size-and-age-dependent birth
  processes.
\newblock {\em\protect\JournalTitle{Journal of Applied Probability}}
  \textbf{11}, 248--254 (1974).

\bibitem{VoitDick1983}
EO Voit, G Dick, Growth of cell populations with arbitrarily distributed cycle
  durations. {I.} {B}asic model.
\newblock {\em\protect\JournalTitle{Mathematical Biosciences}} \textbf{66},
  229--246 (1983).

\bibitem{HirschEngelberg1966}
HR Hirsch, J Engelberg, Decay of cell synchronization: Solutions of the
  cell-growth equation.
\newblock {\em\protect\JournalTitle{The Bulletin of Mathematical Biophysics}}
  \textbf{28}, 391--409 (1966).

\bibitem{Hinshelwood1952}
CN Hinshelwood, On the chemical kinetics of autosynthetic systems.
\newblock {\em\protect\JournalTitle{Journal of the Chemical Society}} pp.
  745--755 (1952).

\bibitem{ScottHwa2011}
M Scott, T Hwa, Bacterial growth laws and their applications.
\newblock {\em\protect\JournalTitle{Current Opinion in Biotechnology}}
  \textbf{22}, 559--565 (2011).

\bibitem{Pandey2020}
PP Pandey, H Singh, S Jain, Exponential trajectories, cell size fluctuations,
  and the adder property in bacteria follow from simple chemical dynamics and
  division control.
\newblock {\em\protect\JournalTitle{Physical Review E}} \textbf{101}, 062406
  (2020).

\bibitem{WangJun2010}
P Wang, et~al., Robust growth of {E}scherichia coli.
\newblock {\em\protect\JournalTitle{Current Biology}} \textbf{20}, 1099--1103
  (2010).

\bibitem{TanouchiYou2017}
Y Tanouchi, et~al., Long-term growth data of {E}scherichia coli at a
  single-cell level.
\newblock {\em\protect\JournalTitle{Scientific Data}} \textbf{4}, 170036
  (2017).

\bibitem{IyerBiswas2014}
S Iyer-Biswas, GE Crooks, NF Scherer, AR Dinner, Universality in stochastic
  exponential growth.
\newblock {\em\protect\JournalTitle{Physical Review Letters}} \textbf{113},
  028101 (2014).

\bibitem{IyerBiswas2014b}
S Iyer-Biswas, et~al., Scaling laws governing stochastic growth and division of
  single bacterial cells.
\newblock {\em\protect\JournalTitle{Proceedings of the National Academy of
  Sciences}} \textbf{111}, 15912--15917 (2014).

\bibitem{HoAmir2018}
PY Ho, J Lin, A Amir, Modeling cell size regulation: From single-cell-level
  statistics to molecular mechanisms and population-level effects.
\newblock {\em\protect\JournalTitle{Annual Review of Biophysics}} \textbf{47},
  251--271 (2018) PMID: 29517919.

\bibitem{Allen1985}
JC Allen, G Joseph, Deterioration of pasteurized milk on storage.
\newblock {\em\protect\JournalTitle{Journal of Dairy Research}} \textbf{52},
  469--487 (1985).

\bibitem{Simon2001}
M Simon, A Hansen, Effect of various dairy packaging materials on the shelf
  life and flavor of ultrapasteurized milk.
\newblock {\em\protect\JournalTitle{Journal of Dairy Science}} \textbf{84},
  784--791 (2001).

\bibitem{Petrus2010}
R Petrus, C Loiola, C Oliveira, Microbiological shelf life of pasteurized milk
  in bottle and pouch.
\newblock {\em\protect\JournalTitle{Journal of Food Science}} \textbf{75},
  M36--M40 (2010).

\bibitem{Membre2004}
JM Membr{\'e}, M Kubaczka, J Dubois, C Ch{\`e}n{\'e}, Temperature effect on
  {L}isteria monocytogenes growth in the event of contamination of cooked pork
  products.
\newblock {\em\protect\JournalTitle{Journal of Food Protection}} \textbf{67},
  463--469 (2004).

\bibitem{PasteurizedMilkOrdinance}
{Food and Drug Administration}, {\em Grade ``A" Pasteurized Milk Ordinance
  229}.
\newblock (US Department of Health and Human Services, Public Health Service,
  Food and Drug Administration), (1995).

\bibitem{Boor2001}
KJ Boor, Fluid dairy product quality and safety: looking to the future.
\newblock {\em\protect\JournalTitle{Journal of Dairy Science}} \textbf{84},
  1--11 (2001).

\bibitem{Feller1939}
W Feller, Die grundlagen der volterraschen theorie des kampfes ums dasein in
  wahrscheinlichkeitstheoretischer behandlung.
\newblock {\em\protect\JournalTitle{Acta Biotheoretica}} \textbf{5}, 11--40
  (1939).

\bibitem{Feller1968}
W Feller, {\em An Introduction to Probability Theory and Its Applications}.
\newblock (Wiley) Vol.{}~1, (1968).

\bibitem{RichterDyn1972}
N Richter-Dyn, NS Goel, On the extinction of a colonizing species.
\newblock {\em\protect\JournalTitle{Theoretical Population Biology}}
  \textbf{3}, 406--433 (1972).

\bibitem{Redner2001}
S Redner, {\em A Guide to First-Passage Processes}.
\newblock (Cambridge University Press), (2001).

\bibitem{Oehlert1992}
GW Oehlert, A note on the delta method.
\newblock {\em\protect\JournalTitle{The American Statistician}} \textbf{46},
  27--29 (1992).

\bibitem{VerHoef2012}
JMV Hoef, Who invented the delta method?
\newblock {\em\protect\JournalTitle{The American Statistician}} \textbf{66},
  124--127 (2012).

\bibitem{BellmanHarris1948}
R Bellman, TE Harris, On the theory of age-dependent stochastic branching
  processes.
\newblock {\em\protect\JournalTitle{Proceedings of the National Academy of
  Sciences}} \textbf{34}, 601--604 (1948).

\bibitem{RolfeHinton2012}
MD Rolfe, et~al., Lag phase is a distinct growth phase that prepares bacteria
  for exponential growth and involves transient metal accumulation.
\newblock {\em\protect\JournalTitle{Journal of Bacteriology}} \textbf{194},
  686--701 (2012).

\bibitem{Bertrand2019}
RL Bertrand, Lag phase is a dynamic, organized, adaptive, and evolvable period
  that prepares bacteria for cell division.
\newblock {\em\protect\JournalTitle{Journal of Bacteriology}} \textbf{201},
  10.1128/jb.00697--18 (2019).

\bibitem{MorenoAckermann2020}
S Moreno-G{\'a}mez, et~al., Wide lag time distributions break a trade-off
  between reproduction and survival in bacteria.
\newblock {\em\protect\JournalTitle{Proceedings of the National Academy of
  Sciences}} \textbf{117}, 18729--18736 (2020).

\bibitem{Hartmann2019}
R Hartmann, et~al., Emergence of three-dimensional order and structure in
  growing biofilms.
\newblock {\em\protect\JournalTitle{Nature Physics}} \textbf{15}, 251--256
  (2019).

\bibitem{Welker2021}
A Welker, et~al., Spatiotemporal dynamics of growth and death within spherical
  bacterial colonies.
\newblock {\em\protect\JournalTitle{Biophysical Journal}} \textbf{120},
  3418--3428 (2021).

\bibitem{Nisbet1986}
R Nisbet, W Gurney, The formulation of age-structure models.
\newblock {\em\protect\JournalTitle{Mathematical Ecology: An Introduction}} pp.
  95--115 (1986).

\bibitem{LinAmir2017}
J Lin, A Amir, The effects of stochasticity at the single-cell level and cell
  size control on the population growth.
\newblock {\em\protect\JournalTitle{Cell Systems}} \textbf{5}, 358--367.e4
  (2017).

\bibitem{Jafarpour2019}
F Jafarpour, Cell size regulation induces sustained oscillations in the
  population growth rate.
\newblock {\em\protect\JournalTitle{Phys. Rev. Lett.}} \textbf{122}, 118101
  (2019).

\bibitem{Nisbet2003}
RM Nisbet, W Gurney, {\em Modelling fluctuating populations: reprint of First
  Edition (1982)}.
\newblock (Blackburn Press), (2003).

\bibitem{Hell2004}
SW Hell, M Dyba, S Jakobs, Concepts for nanoscale resolution in fluorescence
  microscopy.
\newblock {\em\protect\JournalTitle{Current Opinion in Neurobiology}}
  \textbf{14}, 599--609 (2004).

\bibitem{Bar2020}
J B{\"a}r, M Boumasmoud, RD Kouyos, AS Zinkernagel, C Vulin, Efficient
  microbial colony growth dynamics quantification with coltapp, an automated
  image analysis application.
\newblock {\em\protect\JournalTitle{Scientific Reports}} \textbf{10}, 16084
  (2020).

\bibitem{JonesLudington2022}
EW Jones, JM Carlson, DA Sivak, WB Ludington, Stochastic microbiome assembly
  depends on context.
\newblock {\em\protect\JournalTitle{Proceedings of the National Academy of
  Sciences}} \textbf{119}, e2115877119 (2022).

\bibitem{VardakasFalagas2009}
KZ Vardakas, DK Matthaiou, ME Falagas, Incidence, characteristics and outcomes
  of patients with severe community acquired-mrsa pneumonia.
\newblock {\em\protect\JournalTitle{European Respiratory Journal}} \textbf{34},
  1148--1158 (2009).

\bibitem{Mira2022}
P Mira, P Yeh, BG Hall, Estimating microbial population data from optical
  density.
\newblock {\em\protect\JournalTitle{PLOS ONE}} \textbf{17}, e0276040 (2022).

\bibitem{Gaudy1963}
A Gaudy{,}~Jr, F Abu-Niaaj, E Gaudy, Statistical study of the spot-plate
  technique for viable-cell counts.
\newblock {\em\protect\JournalTitle{Applied Microbiology}} \textbf{11},
  305--309 (1963).

\bibitem{StevensonPilizota2016}
K Stevenson, AF McVey, IBN Clark, PS Swain, T Pilizota, General calibration of
  microbial growth in microplate readers.
\newblock {\em\protect\JournalTitle{Scientific Reports}} \textbf{6}, 38828
  (2016).

\bibitem{Knuth1997}
DE Knuth, {\em The Art of Computer Programming, Volume 1}.
\newblock (Bulletin of the American Mathematical Society), (1997).

\bibitem{Keyfitz1997}
B Keyfitz, N Keyfitz, The mckendrick partial differential equation and its uses
  in epidemiology and population study.
\newblock {\em\protect\JournalTitle{Mathematical and Computer Modelling}}
  \textbf{26}, 1--9 (1997).

\bibitem{Higgins1997}
K Higgins, A Hastings, JN Sarvela, LW Botsford, Stochastic dynamics and
  deterministic skeletons: population behavior of dungeness crab.
\newblock {\em\protect\JournalTitle{Science}} \textbf{276}, 1431--1435 (1997).

\end{thebibliography}

\end{document}